\documentclass[twocolumn]{aastex631}
\usepackage{amsmath}
\usepackage{xcolor}
\usepackage{footmisc}
\usepackage{rotating,longtable}
\usepackage{tablefootnote}
\usepackage{booktabs}
\usepackage{comment}

\usepackage{graphicx}
\usepackage{natbib} 

\usepackage{tablefootnote}
\usepackage{booktabs}

\usepackage{lipsum,mwe}
\usepackage{hyperref}
\newcommand{\kms}{\,km\,s$^{-1}$}
\newcommand{\tco}{$^{13}$CO\,(1-0)}
\newcommand{\cetno}{C$^{18}$O\,(1-0)}
\newcommand{\co}{$^{12}$CO\,(1-0)}
\newcommand{\halpha}{H$\alpha$}
\received{---}
\accepted{---}
\shorttitle{[HKS2019] E71}
\shortauthors{Verma et. al.}
\begin{document}

\title{Unraveling the Feedback-Regulated Star Formation Activities around the Expanding Galactic MIR Bubble [HKS2019] E71}

\correspondingauthor{Aayushi Verma}
\email{aayushiverma@aries.res.in}
% \email{vermaaayushi16@gmail.com}

\author[0000-0002-6586-936X]{Aayushi Verma}
\affil{Aryabhatta Research Institute of observational sciencES (ARIES),
Manora Peak, Nainital 263001, India}
\affil{Department of Applied Physics / Physics, M.J.P.Rohilkhand University, Bareilly, Uttar Pradesh-243006, India}

\author[0000-0001-5731-3057]{Saurabh Sharma}
\affil{Aryabhatta Research Institute of observational sciencES (ARIES),
Manora Peak, Nainital 263001, India}

\author[0000-0001-6725-0483]{Lokesh K. Dewangan}
\affil{Physical Research Laboratory, Navrangpura, Ahmedabad - 380009, India}

\author[0009-0008-8490-8601]{Tarak Chand}
\affil{Aryabhatta Research Institute of observational sciencES (ARIES),
Manora Peak, Nainital 263001, India}
\affil{Department of Applied Physics / Physics, M.J.P.Rohilkhand University, Bareilly, Uttar Pradesh-243006, India}

\author[0009-0003-6633-525X]{Ariful Hoque}
\affil{S. N. Bose National Centre for Basic Sciences, Sector III, Salt Lake, Kolkata-700106, India}

\author[0000-0001-9312-3816]{Devendra K. Ojha} 
\affil{Tata Institute of Fundamental Research (TIFR), Homi Bhabha Road, Colaba, Mumbai - 400005, India}

\author[0000-0002-0444-0439]{Harmeen Kaur}
\affil{Aryabhatta Research Institute of observational sciencES (ARIES),
Manora Peak, Nainital 263001, India}

\author[0000-0002-6740-7425]{Ram Kesh Yadav}
\affiliation{National Astronomical Research Institute of Thailand (Public Organization), 260 Moo 4, T. Donkaew, A. Maerim, Chiangmai 50 180, Thailand}

\author[0009-0001-4144-2281]{Mamta}
\affil{Aryabhatta Research Institute of observational sciencES (ARIES),
Manora Peak, Nainital 263001, India}

\author[0009-0009-6420-8058]{Manojit Chakraborty}
\affil{Aryabhatta Research Institute of observational sciencES (ARIES),
Manora Peak, Nainital 263001, India}

\author{Archana Gupta}
\affil{Department of Applied Physics / Physics, M.J.P.Rohilkhand University, Bareilly, Uttar Pradesh-243006, India}

\begin{abstract}
We explore the physical environment of the Galactic mid-infrared (MIR) bubble [HKS2019]~E71 (hereafter E71) through a multi-wavelength approach. E71 is located at the edge of a filamentary structure, as traced in \textit{Herschel} images (250–500~$\mu$m), \textit{Herschel} column density map, and molecular maps in the velocity range [$-20$,~$-14$] \kms. It hosts a stellar cluster (radius$\sim$1.26 pc, distance$\sim$$1.81\pm0.15$ kpc) associated with radio continuum emission, including a centrally positioned B1.5-type massive star (hereafter `m2'), along with an enhanced population of evolved low-mass stars and young stellar objects.
MIR images and molecular line maps reveal a PDR surrounding `m2', exhibiting an arc-like structure along the edges of E71. Regularly spaced molecular and dust condensations are identified along this structure. The position-velocity map of \co\, emission suggests an expansion of molecular gas concentrated at the periphery of E71. Near-infrared spectroscopic observations with TANSPEC confirm the presence of the accretion process in a massive young stellar object (MYSO) located near the edge of the bubble. High-resolution uGMRT radio continuum maps uncover substructures in the ionized emission, both toward the MYSO and the center of E71. These findings support that `m2' has shaped an arc-like morphology through its feedback processes. The pressure exerted by `m2' and the velocity structure of the $^{12/13}$CO\,(1-0) emission suggest that the stellar feedback has likely driven out molecular material, leading to the formation of the expanding E71 bubble. Our overall investigation infers that the ``collect and collapse'' process might be a possible mechanism that can describe the ongoing star formation activities around the E71 bubble.
% This swept-up gas is now hosting active star formation. 

%\textbf{We also infer that the MYSO likely formed as a direct result of the radiative and mechanical feedback from the massive B1.5-type star.}

\end{abstract}

%% keywords and the rules for their use.(ISM:) H\,{\sc ii} regions



%%%%%%%%%%%%%%%%% BODY OF PAPER %%%%%%%%%%%%%%%%%%

\section{Introduction} \label{sec:intro}

Large-scale infrared (IR) surveys (e.g., \citealt{2006ApJ...649..759C, 2010AJ....139.2330W, 2019PASJ...71....6H}) have revealed some of the most striking structures in molecular clouds: IR ‘bubbles’. These shell- or arc-like features are formed when expanding H\,{\sc ii} regions, driven by the radiative and mechanical feedback from massive star(s) (with masses $\geq8\,M_\odot$), interact with the surrounding molecular material, leading to the formation of photo-dissociation regions (PDRs). The expansion of such bubbles can induce massive star formation along their peripheries \citep{2012A&A...544A..11Z, 2013MNRAS.431.1062D}, as the neutral gas is compressed, potentially becoming gravitationally unstable, fragmenting into dense cores, and forming a new generation of massive stars \citep{1977ApJ...214..725E, 2010A&A...523A...6D, 2020ApJ...898...41D, 2020A&A...642A..87K, 2020ApJ...897...74Z}. Therefore, studying these IR bubbles offers crucial insights into the interaction between H\,{\sc ii} regions and their host molecular clouds, as well as into the mechanisms of triggered star formation.

 \begin{figure*}[!ht]
    \centering
    \includegraphics[width=0.49\textwidth]{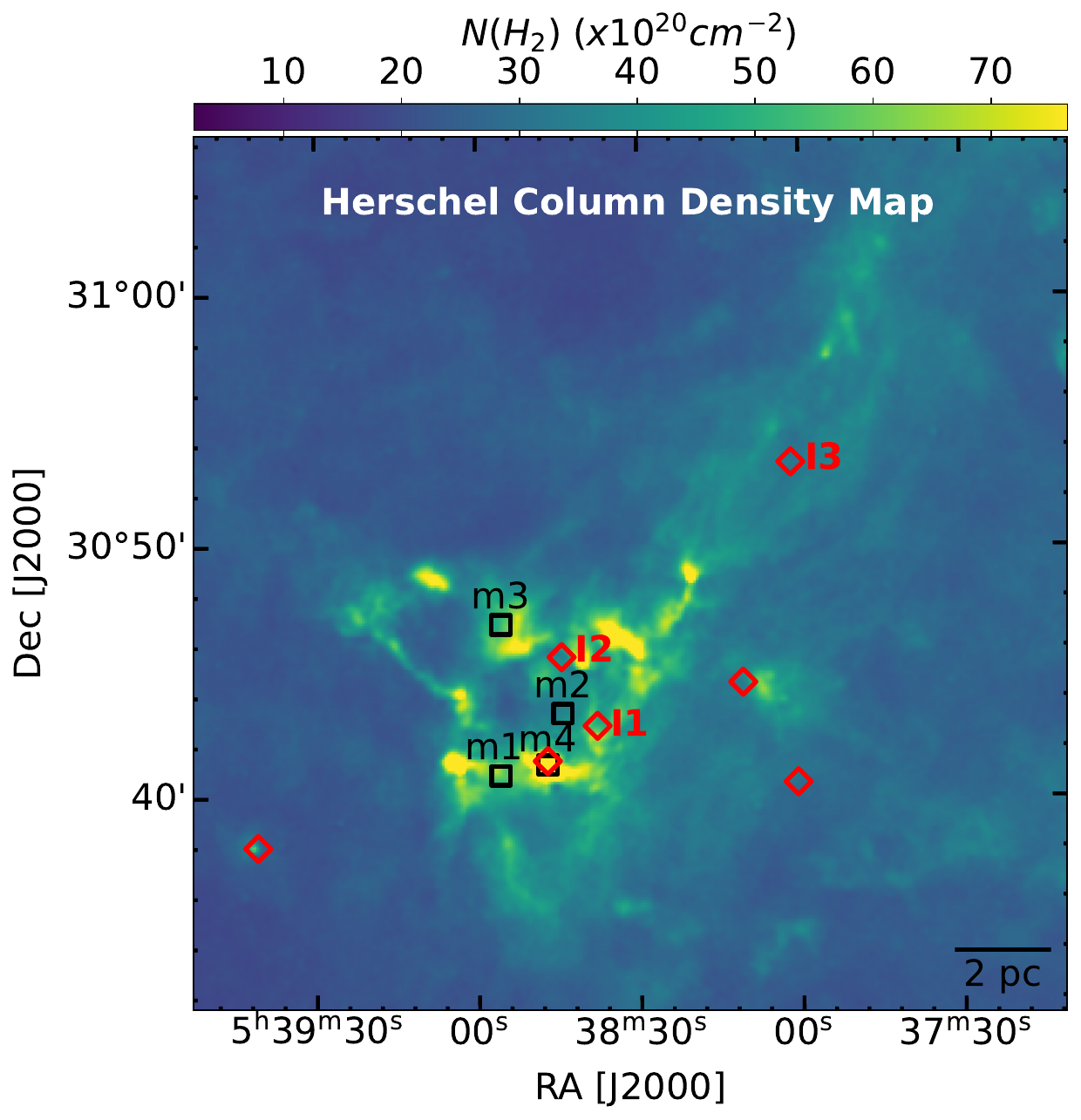}
    \includegraphics[width=0.49\textwidth]{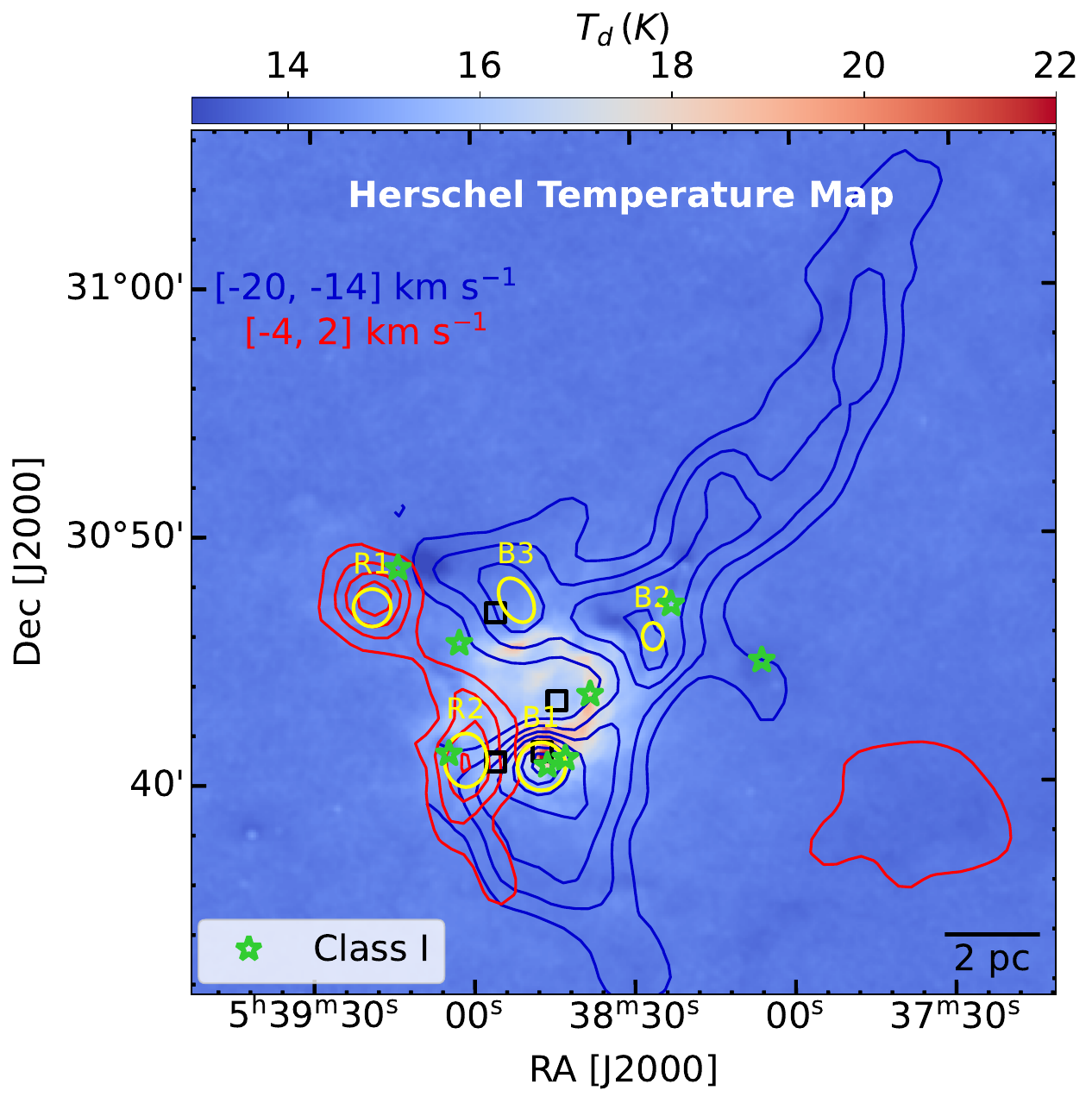}
    \includegraphics[width=0.98\textwidth]{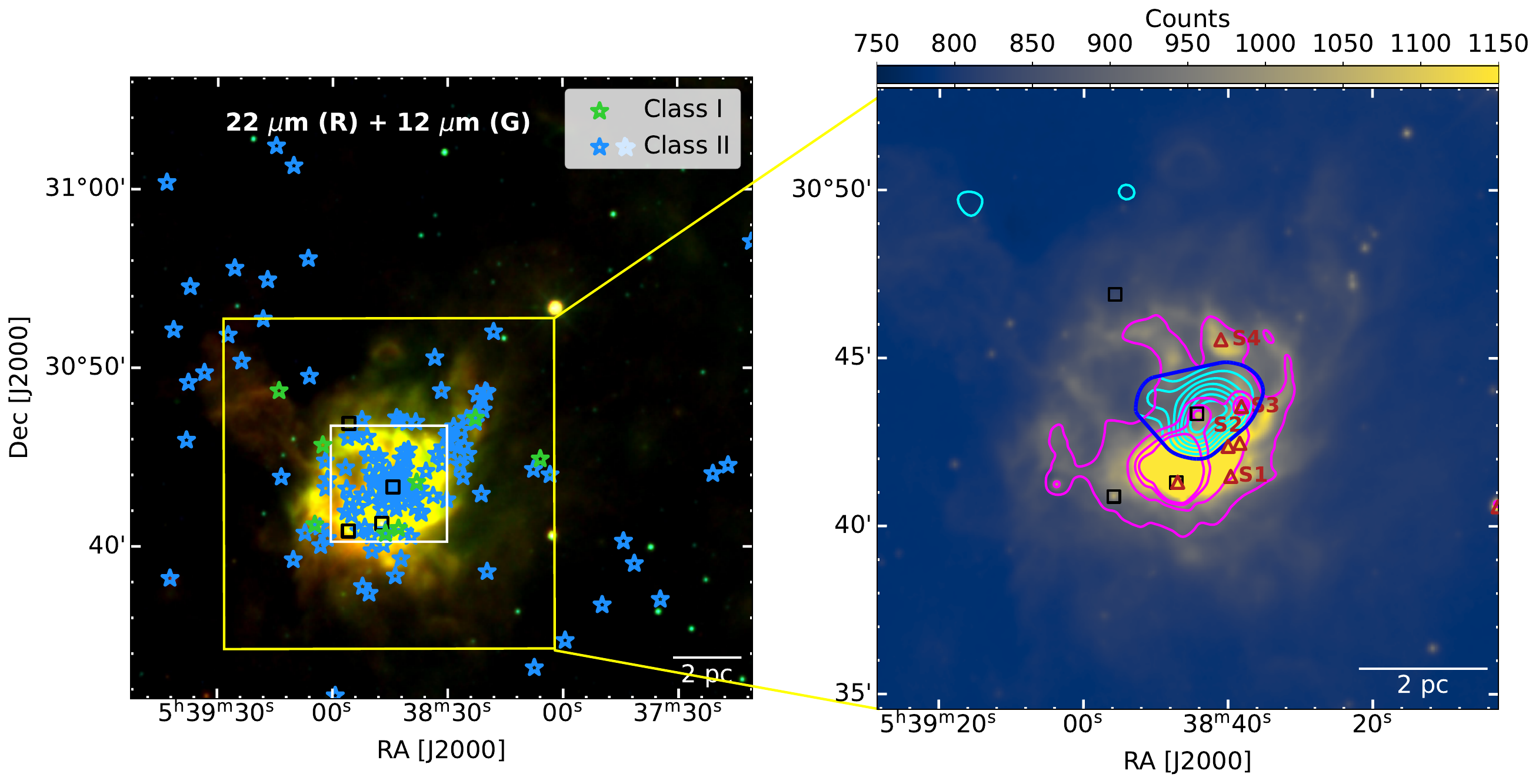}
    \caption{Upper left panel: \textit{Herschel} column density map showing the large-scale view ($35\arcmin \times 35\arcmin$) of the E71 bubble, overlaid with the locations of IRAS sources (red diamonds). Upper right panel: \textit{Herschel} dust temperature map overlaid with the locations of Class\,\textsc{i} YSOs, and blue and red contours representing $^{12}$CO integrated intensity in the velocity ranges [$-20$, $-14$] km s$^{-1}$ and [$-4$, $2$] km s$^{-1}$, respectively (Section~\ref{sec:ChannelMap}). The lowest contour levels correspond to emission above the 5$\sigma$ threshold, where $\sigma$ is the rms noise. This map is also overlaid with the molecular condensations, marked with yellow circles/ellipses. Lower left panel: Color-composite image generated using WISE 22~$\mu$m (red) and WISE 12~$\mu$m (green) emission. The image is overlaid with yellow and white squares representing the $18\farcm5 \times 18\farcm5$ and $6\farcm5 \times 6\farcm5$ fields observed with the 1.3-m Devasthal Fast Optical Telescope (DFOT) and the 3.6-m Devasthal Optical Telescope (DOT), respectively. The positions of identified Class\,\textsc{i} and Class\,\textsc{ii} YSOs (Section~\ref{app:yso_identification}) are shown as green and blue asterisks, respectively. Lower right panel: Zoomed-in view of the WISE 12~$\mu$m emission of the lower left panel overlaid with the WISE 22~$\mu$m emission (magenta contours), with the lowest contour at 250 counts with a step size of 8.75 counts. Additionally, this panel is overlaid with the locations of MSX sources (red triangles) and near-infrared (NIR) stellar surface density contours (cyan) derived from the UKIDSS-2MASS catalog (see Section~\ref{sec:clustering}), enclosed by a blue polygon representing the \textit{convex hull}. The lowest isodensity contour level is 6.36 stars arcmin$^{-2}$ with a step size of 1.14 stars arcmin$^{-2}$. The locations of candidate massive stars `m1, m2, m3, and m4' are marked with black squares in all panels. }
    \label{fig:intro}
\end{figure*}

This study focuses on the Galactic mid-infrared (MIR) bubble [HKS2019] E71 (hereafter E71; $\alpha_{J2000}=05^h38^m45^{s}.4$, $\delta_{J2000}=+30^{\circ}43\arcmin31\arcsec$), which was first cataloged by \citet{2019PASJ...71....6H} using AKARI and \textit{Herschel} photometric data and has not been extensively explored to date.
The upper-left panel of Figure~\ref{fig:intro} shows the \textit{Herschel} column density map (see Section \ref{sec:archival_data}), offering a large-scale view of the target region. A partial ring (or arc) of gas and dust is evident on the western side with several dense clumps. The IRAS sources present in this region are also marked with red diamonds. The corresponding dust temperature map (upper-right panel; see Section \ref{sec:archival_data}) reveals elevated dust temperatures (i.e., $T_d \sim 18$ K) along the arc, suggesting radiative heating from embedded massive star(s). Overlaid blue and red contours denote $^{12}$CO integrated intensities in the velocity ranges [$-20$, $-14$] km s$^{-1}$ and [$-4$, $2$] km s$^{-1}$, respectively (see Section~\ref{sec:ChannelMap}), indicating the presence of two kinematically distinct molecular clouds in the region. This panel also includes an overlay of Class\,{\sc i} YSOs to examine their spatial distribution relative to the molecular gas. A clear spatial correlation is evident, indicating that the region is actively forming stars. The lower-left panel presents a WISE color-composite image (22 $\mu$m in red; 12 $\mu$m in green). The 22 $\mu$m emission traces warm dust heated by stellar feedback, while the 12 $\mu$m band captures polycyclic aromatic hydrocarbon (PAH) emission at 11.3 $\mu$m. The lower-right panel shows a zoomed-in view of the WISE 12 $\mu$m emission map overlaid with WISE 22 $\mu$m emission contours (magenta). A partial PAH ring, characteristic of a PDR, is more prominently visible in this zoomed-in view, consistent with feedback-driven shell morphology \citep{2016MNRAS.461.2502Y,2020ApJ...896...29K, 2020ApJ...905...61P}. 
Stellar surface density contours (cyan), derived from the United Kingdom Infra-Red Telescope (UKIRT) Infrared Deep Sky Survey and Two Micron All-Sky Survey's combined catalog (UKIDSS–2MASS catalog; see Section~\ref{sec:clustering}), are overlaid in the zoomed-in panel, with a blue \textit{convex hull} outlining a stellar cluster enclosed within the E71 bubble (see Section~\ref{sec:clustering}). MSX sources are marked in this panel with red triangles.
The region also hosts a massive young stellar object (MYSO), labeled ‘m4’ ($\alpha_{J2000}=05^h38^m47^{s}.16$, $\delta_{J2000}=+30^{\circ}41\arcmin18\arcsec$.1), identified in the right panel of Figure~\ref{fig:intro} \citep{2011MNRAS.418.1689U, 2013yCat..22080011L, 2021MNRAS.504..338P}. MYSOs represent an early evolutionary phase of massive OB-type stars and are key to understanding massive star formation processes. They drive strong stellar winds \citep{1995MNRAS.272..346B} and bipolar molecular outflows \citep{1996ApJ...472..225S}, contributing to significant feedback in their natal environments. This MYSO, located at a distance of $2.0 \pm 0.6$ kpc \citep{1998ApJS..117..387K}, shows He \textsc{i} and Br$\gamma$ lines with a P-Cygni profile, indicative of strong outflows \citep{2013MNRAS.430.1125C}. Very Large Array (VLA) observations at 5.8 GHz further classify it as a jet candidate \citep{2021MNRAS.504..338P}. \citet{2011MNRAS.418.1689U}  detected the $\rm NH_3$~(1,1) and $\rm NH_3$~(2,2) emission with the systematic velocity of $-16.14$~\kms\, and $-16.09$~\kms, respectively, towards this MYSO. These velocities closely match with the average velocity ($\sim -16$~\kms) of the molecular cloud associated with the E71 bubble (see Section \ref{sec:ChannelMap}). These features strongly suggest a physical association of the MYSO with the E71 bubble.

% Its location along the E71 arc strongly suggests a physical association with the bubble.

These multi-wavelength features make E71 an ideal star-forming site for examining the influence of massive stellar feedback on the interstellar medium (ISM) and the potential triggering of subsequent star formation.

In this work, we conduct a comprehensive multi-wavelength study of the E71 bubble. The structure of this paper is as follows: Section~\ref{sec:data} describes the observational datasets, reduction methods, and ancillary data used. Section~\ref{sec:result} details the stellar clustering and young stellar objects' (YSOs) population associated with E71 and the characterization of its surrounding physical environment. Section~\ref{sec:discussion} discusses our findings in the context of feedback and triggered star formation, while Section~\ref{sec:conclusion} summarizes our main conclusions.

%%%%%%%%%%%%%%%%%%%%%%%%%%%%%%%%%%%%%%%%

\section{Observations and data reductions}\label{sec:data}

\subsection{Optical Photometric Observation and Reduction}

\subsubsection{SDSS $g$ and $i$ bands}

The deep optical photometric observations of the E71 bubble were performed using the 4K$\times$4K CCD IMAGER mounted at the axial port of the 3.6-m  Devasthal Optical Telescope (DOT), Nainital \citep{2018BSRSL..87...29K,2022JApA...43...27K}. These observations were carried out for a $6^\prime.5 \times6^\prime.5$ region (field of view (FOV) of IMAGER; shown with a white square in Figure~\ref{fig:intro}) on 2023 November 20 in SDSS $g$ and $i$ bands along with a few flat and bias frames. The images were taken for a total integration time of 2 hours and 1.5 hours in $g$ and $i$ bands, respectively, in $2\times2$ binning modes. The basic data reduction (image cleaning, photometry, and astrometry) was done using the standard procedure explained in \citet{2020MNRAS.498.2309S}. A color-composite image of the $i$ (red) and $g$  (green) band images obtained by IMAGER is shown in the upper panel of Figure~\ref{fig:optical_mag}. The instrumental magnitudes were then calibrated using PS1 DR2 data (see Table~\ref{tab:archival_data}). The obtained calibration equations are as follows:

\begin{equation}
    \begin{split}
        (g-i)_{std}& = (0.9542 \pm 0.0237) \times (g-i)_{obs} \\
        &+ 0.1721 \pm 0.0538
    \end{split}
\end{equation}
\begin{equation}
    \begin{split}
        (g_{std}-g_{obs})& = (0.0060 \pm 0.0269) \times (g-i)_{std} \\
        &+ 5.1069 \pm 0.0628
    \end{split}
\end{equation}
where, $g_{std},\,i_{std}$ are the standard magnitudes taken from PS1 whereas $g_{obs},\,i_{obs}$ are the observed instrumental magnitudes from DOT observations. We achieved a magnitude limit of 24.7 mag in the $g$ band with an error less than 0.1 mag.

\subsubsection{Completeness of the Photometric Data}\label{sec:completeness}

\begin{figure}[!ht]
    \centering
    \includegraphics[width=\linewidth]{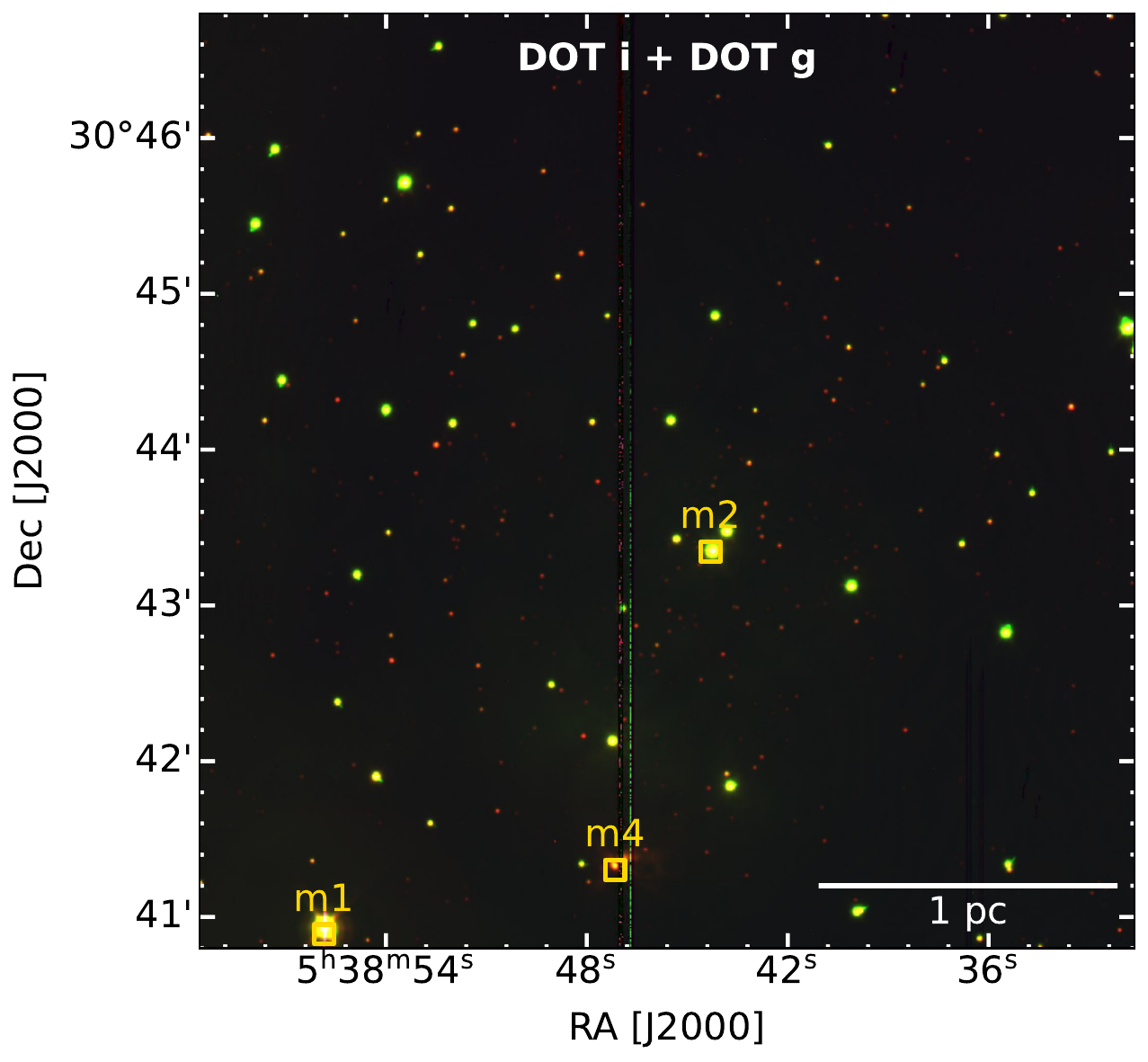 }
    \includegraphics[width=0.9\linewidth]{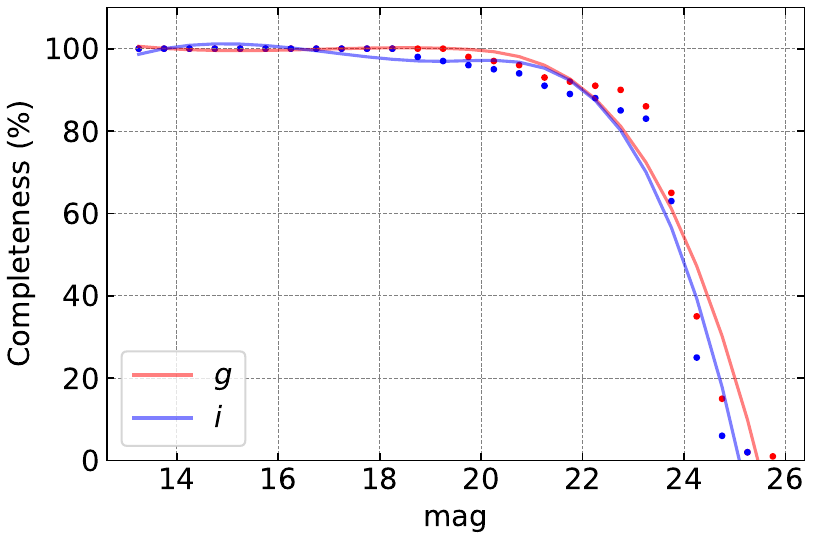}
    \caption{Upper Panel: Color-composite image of the E71 bubble, generated using the $g$ and $i$ band images (green and red, respectively) of $6^\prime.5 \times6^\prime.5$ region (FOV of DOT that entirely covers the bubble). The locations of probable massive stars `m1, m2, and m4' are also marked with yellow squares. Lower Panel: Completeness factor in $g$ and $i$ bands as a function of magnitude derived using the {\sc addstar} routine of IRAF.}
    \label{fig:optical_mag}
\end{figure}

The photometric data might be incomplete due to several factors, such as the instrument's detection limit, crowding of the stars, nebulosity, etc. We estimated the completeness factor (CF) of the photometric data using the {\sc addstar} routine of Image Reduction and Analysis Facility (IRAF; see \citealt{2008AJ....135.1934S} for details). A few fake stars of known magnitude and position are randomly added to the observed frames, which are reduced following the same method as the original frames. CF is then estimated as the ratio of recovered stars to the number of artificially added stars in the image frame, in different magnitude bins.  As anticipated, the resultant completeness of the photometric data decreases as we go to the fainter magnitudes (see lower panel of Figure~\ref{fig:optical_mag}). 
The stars of $\sim$24 magnitudes were detected in $g$ and $i$ bands with a $\ge 50\%$ CF. 

%%%%%%%%%%%%%%%%%%%%%%%%%%%%%%%%%%%%%%%%

\subsubsection{Narrow-band \halpha\, and Johnson $R$ band} \label{sec:halpha}
We also observed the E71 bubble using the 1.3-m Devasthal Fast Optical Telescope (DFOT), Nainital, on 2023 November 20, in narrow-band \halpha\, and Johnson $R$ band using a 2K $\times$ 2K CCD camera having a FOV of $18\arcmin.5 \times18\arcmin.5$ (\citealt{2012ASInC...4..173S}, shown with a yellow square in Figure~\ref{fig:intro}). Several flat and bias frames were also taken during the observations. The image cleaning was done using the standard procedure explained in \citet{2020MNRAS.498.2309S}. The $R$ band image was scaled to the \halpha\, image after accounting for the FWHM difference of the stellar profiles and then was subtracted from it to get the \halpha\, line image \citep{2014MNRAS.439..157K}.

%%%%%%%%%%%%%%%%%%%%%%%%%%%%%%%%%%%%%%%%

\subsection{Optical Spectroscopic Observation and Reduction} \label{sec:hct_spectra}

\begin{figure}[!ht]
    \centering
    \includegraphics[width=0.88\linewidth]{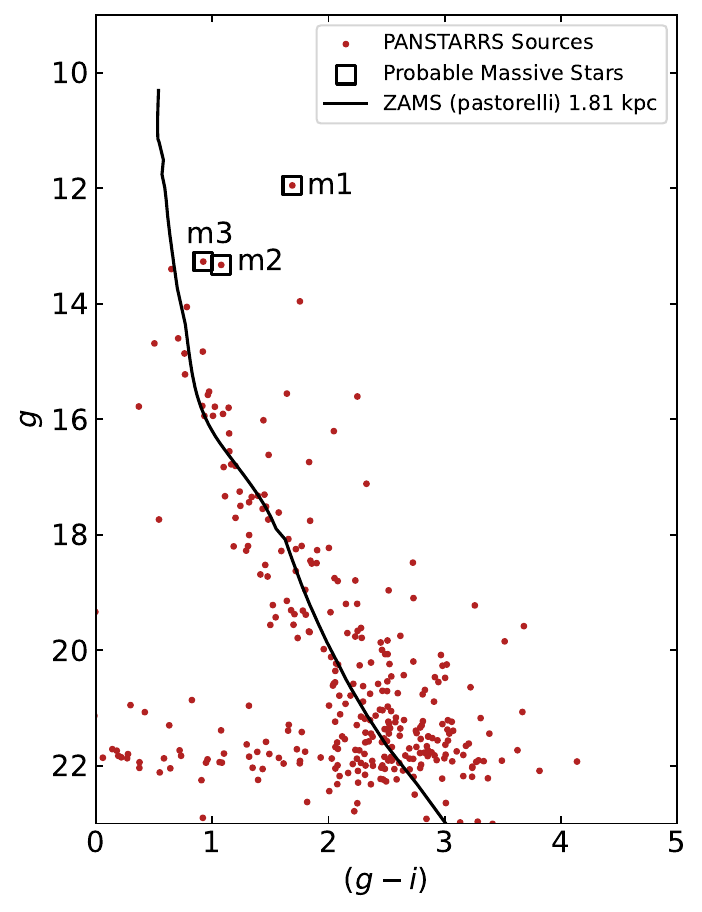 }
    \caption{$g$ versus $(g-i)$ CMD for the optical sources detected by PS1 DR2 data. The black curve represents the ZAMS isochrones by \citet{2019MNRAS.485.5666P}, corrected for the distance 1.81 kpc and $A_V$ =2.63 mag. The probable massive stars (`m1, m2, and m3') are marked with black squares.}
    \label{fig:panstarrs_cmd}
\end{figure}

Given that this region exhibits typical H\,{\sc ii} region characteristics and clear signs of stellar feedback, we searched for massive star candidates based on their positions in the optical Hertzsprung–Russell (HR) diagram or the color-magnitude diagram (CMD; \citealt{2024AJ....167..106S,2006AJ....132.1669S}), using PAN-STARRS1 (PS1) DR2 photometric data (see Figure \ref{fig:panstarrs_cmd}; \citealt{2016arXiv161205560C}). Three such candidates were identified, labeled as `m1', `m2', and `m3' (coordinates are mentioned in Table \ref{tab:massive_ra_dec}) and marked in all the panels of Figure~\ref{fig:intro}. Follow-up spectroscopic observations of these sources were conducted using the Hanle Faint Object Spectrograph Camera (HFOSC) on the 2.0-m Himalayan Chandra Telescope (HCT), Hanle, India, on 2023 December 04. Observations were made with GRISM 7 (wavelength range 3800–6840 \r{A}, resolving power R$\sim$1200). For wavelength calibration, an FeAr calibration lamp was observed on the same night.

\begin{table}[!ht]
    \centering
    \caption{Coordinates of the probable massive stars}
    \begin{tabular}{c c c}
    \hline
    Name & $\alpha_{J2000}$ & $\delta_{J2000}$\\
     & (hh:mm:ss.ss) & (dd:mm:ss.ss)\\
    \hline
    m1 & 05:38:55.85 & +30:40:53.37\\
    m2 & 05:38:44.30 & +30:43:20.08\\
    m3 & 05:38:55.70 & +30:46:54.02\\
    \hline\\
    \end{tabular}
   
    \label{tab:massive_ra_dec}
\end{table}

The data were reduced with IRAF packages following the procedure illustrated in \citet{2012MNRAS.424.2486J}. Aperture extraction, line identification by lamp, and dispersion correction were achieved by {\sc apall, identify}, and {\sc dispcor} tasks, respectively. And finally, the normalized wavelength-calibrated spectra were achieved using the {\sc continuum} task. 

%%%%%%%%%%%%%%%%%%%%%%%%%%%%%%%%%%%%%%%%

\subsection{Optical-NIR Spectroscopic Observation and Reduction} \label{sec:tanspec_spectra}

We carried out the spectroscopic observations of the MYSO `m4' using the TIFR-ARIES Near Infrared Spectrometer (TANSPEC) mounted at the main port of 3.6-m DOT \citep{2022PASP..134h5002S}. It observes in two spectroscopic modes: (a) relatively high-resolution/ medium-resolution cross-dispersed (XD) mode and (b) a lower resolution prism mode, covering from optical (0.55 $\mu$m) to near-infrared (NIR; 2.5 $\mu$m) bands. We took the spectra of `m4' in XD mode with 1$\arcsec$ slit, which serves a median resolution (R) of about 1500. We followed the standard reduction procedures to obtain the spectrum of `m4'. We used the {\sc apall} task from IRAF to extract spectra from multiple spectral orders in the 2D spectral image. Wavelength calibration was then performed using the module available in \textit{pyTANSPEC} pipeline \citep{2023JApA...44...50G}.
% We also observed a telluric standard star HD71906 of the spectral type A0V for telluric correction along with the Neon and Argon lamps for flat fielding.
We have normalized the output spectrum using the {\sc continuum} task for different orders and then combined them to generate the final spectrum.

%%%%%%%%%%%%%%%%%%%%%%%%%%%%%%%%%%%%%%%%

\subsection{Molecular line data}
\label{sec:MolecularCOData}
We utilized the molecular \co, \tco, and \cetno\ line data observed by Purple Mountain Observatory (PMO) 13.7\,m millimeter-wave radio telescope as a part of the Milky Way Imaging Scroll Painting (MWISP) project \citep{ 633694461037117445,Su_MWISP_2019ApJS}. The channel width for \co, \tco, and \cetno\, spectral data cubes are 0.16, 0.17, and 0.17 \kms, respectively, and the grid spacing and the beam size are $\sim$30\arcsec\, and $\sim$50\arcsec, respectively. The rms noises were estimated to be $\sim$\,0.51\,K ($^{12}$CO), $\sim$\,0.24\,K ($^{13}$CO), and $\sim$\,0.25\,K (C$^{18}$O), with the pixel brightness expressed in T$_{\mathrm{MB}}$ (main beam temperature) units. Out of \co, \tco, and \cetno; \co\, is the most consistent to reveal the spatial extent of the diffused gas (with a gas density of $\sim$10$^2$ cm$^{-3}$) due to its enormous abundance \citep{Su_MWISP_2019ApJS}. Since we did not observe any strong emission in \cetno\, transition, we have not included that in our study. The further study is restrained above the 5$\sigma$ threshold emission ($\sigma$ being the rms noise) to get the fundamental physical features and eliminate any baffling artifact.

%%%%%%%%%%%%%%%%%%%%%%%%%%%%%%%%%%%%%%%%

\subsection{Radio Continuum Data}

We carried out the radio interferometric observations at 
% \textbf{610 (Band 4) and} 
1260 MHz (Band 5) of the E71 bubble using the upgraded Giant Metrewave Radio Telescope (uGMRT; \citealt{SwarupGMRT_91,Gupta_ugmrt_2017CSci}) on 
% \textbf{2023 October 31} and 
2023 November 01 (PI: Aayushi Verma; Proposal ID: 45\_074). uGMRT provides enhanced sensitivity for continuum imaging and exquisite $uv$ coverage to map diffuse, extended emission. For our analysis, we utilized the GMRT Wideband Backend (GWB) data, corresponding to a bandwidth of 200 MHz. To authorize the flux density scale, we observe the flux calibrator 3C147 at the beginning and the end of the observation, and the phase calibrator 0555$+$398 in between while observing the target source E71.

The data was reduced using the Common Astronomy Software Applications (CASA) software \citep{2022PASP..134k4501C}, including the following steps: several iterations of ``flagging'' the corrupt data due to radio frequency interference (RFI) using {\sc flagdata} task and ``calibration'' (including flux density, delay, bandpass, gain calibration using {\sc setjy}, {\sc gaincal}, {\sc bandpass}, {\sc fluxscale}, and {\sc applycal} tasks), splitting and averaging the calibrated target data using {\sc mstransform} task and imaging with a few rounds of self-calibration using {\sc tclean} task. Finally, the primary beam correction was performed using the task {\sc ugmrtpb}\footnote{\url{https://github.com/ruta-k/uGMRTprimarybeam-CASA6}}, which has been written particularly for the uGMRT. The beam size and the rms noise of the final image were $\sim2.4\arcsec\times2.0\arcsec$ and $\sim5.0~\mu$Jy/beam, respectively.
% ; \textbf{whereas for Band 4, these values were $\sim4.5\arcsec\times3.5\arcsec$ and \textbf{$\sim7.6~\mu$Jy/beam}, respectively.}
%%%%%%%%%%%%%%%%%%%%%%%%%%%%%%%%%%%%%%%%

\subsection{Other Ancillary Data}\label{sec:archival_data}

\begin{table*}[!ht]
    \footnotesize
    \centering
    \caption{Ancillary data sets employed for the current study (optical to radio wavelength regime).}
    \begin{tabular}{c c c c}
    \hline
    Survey & Wavelength/s & $\sim$ Resolution & Reference\\
    \hline
    Pan-STARRS1 Surveys\footnote{https://catalogs.mast.stsci.edu/panstarrs/} (PS1; g and i) & 4866 and 7545 \AA & $0\arcsec.25$ & \citet{2016arXiv161205560C}\\
    \emph{Gaia} DR3\footnote{https://www.cosmos.esa.int/web/gaia/dr3} (magnitudes, parallax, and PM) & 330–1050 nm & 0.4 mas & \citet{2016gaia,2023gaia}\\
    Two Micron All Sky Survey\footnote{\citet{https://doi.org/10.26131/irsa2}} (2MASS) & 1.25, 1.65, and 2.17 $\mu$m & $2\arcsec.5$ & \citet{2006AJ....131.1163S}\\
    UKIRT InfraRed Deep Sky Survey\footnote{http://wsa.roe.ac.uk/} (UKIDSS) & 1.25, 1.65, and 2.22 $\mu$m & $0\arcsec.8$, $0\arcsec.8$, $0\arcsec.8$ & \citet{2008MNRAS.391..136L}\\
    \textit{Spitzer} GLIMPSE360 Survey\footnote{\citet{https://doi.org/10.26131/irsa214}} & 3.6 and 4.5 $\mu$m & $2\arcsec$, $2\arcsec$ & \citet{2005ApJ...630L.149B}\\
    Wide-field Infrared Survey Explorer\footnote{\citet{https://doi.org/10.26131/irsa1}} (WISE) & 3.4, 4.6, 12, 22 $\mu$m & $6\arcsec.1$, $6\arcsec.4$, $6\arcsec.5$, $12\arcsec$ & \citet{2010AJ....140.1868W}\\
    \textit{Herschel} Infrared Galactic Plane Survey\footnote{http://archives.esac.esa.int/hsa/whsa/} & 70, 160, 250, 350, 500 $\mu$m & $5\arcsec.8$, $12\arcsec$, $18\arcsec$, $25\arcsec$, $37\arcsec$ & \citet{2010PASP..122..314M}\\
    NRAO VLA Sky Survey\footnote{https://www.cv.nrao.edu/nvss/postage.shtml} (NVSS) & 21 cm & $45\arcsec$ & \citet{1998AJ....115.1693C}\\
    Milky Way Imaging Scroll Painting (MWISP)  & \co\, and \tco  &  $50\arcsec$ & \citet[]{Su_MWISP_2019ApJS} \\
    \hline
    \end{tabular}
    \label{tab:archival_data}
\end{table*}

We used several ancillary data sets from optical to radio wavelength regimes, concisely specified in Table~\ref{tab:archival_data}. The \textit{Herschel} column density, differential column density, and temperature maps (spatial resolution $\sim$12$\arcsec$) were downloaded directly from the open website\footnote{\url{http://www.astro.cardiff.ac.uk/research/ViaLactea/}}. These maps have been procured for the EU-funded ViaLactea project \citep{2010PASP..122..314M} utilizing the Bayesian Point Process Mapping (PPMAP) technique \citep{2010A&A...518L.100M} at 70, 160, 250, 350, and 500 $\mu$m wavelengths \textit{Herschel} data \citep{2015MNRAS.454.4282M, 2017MNRAS.471.2730M}.

%%%%%%%%%%%%%%%%%%%%%%%%%%%%%%%%%%%%%%%%

\section{Result and Analysis}\label{sec:result}

\subsection{Spectral Analysis of `m2 and m3'}
\begin{figure*}[!ht]
    \centering
    \includegraphics[width=\textwidth]{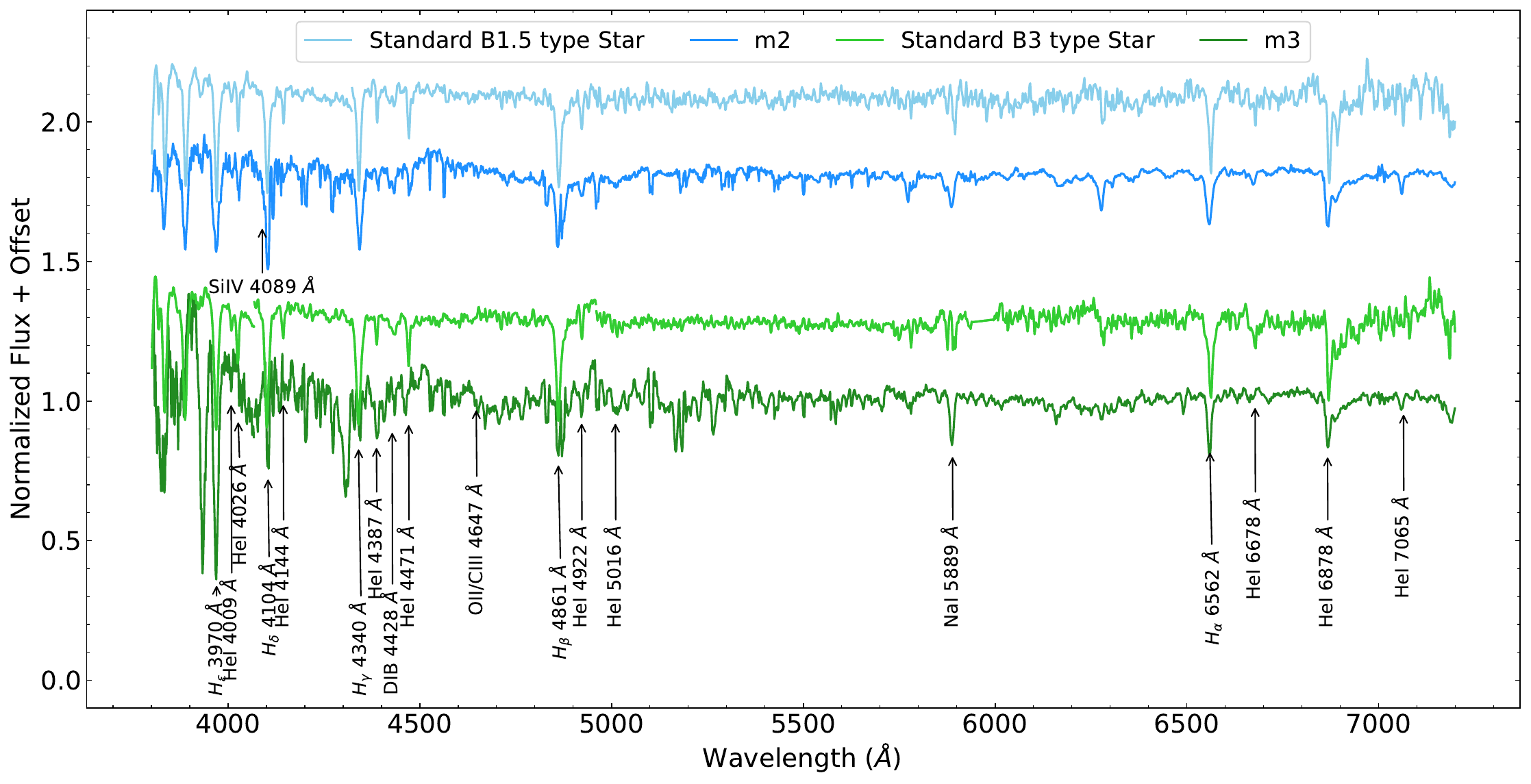}
    \includegraphics[width=\textwidth]{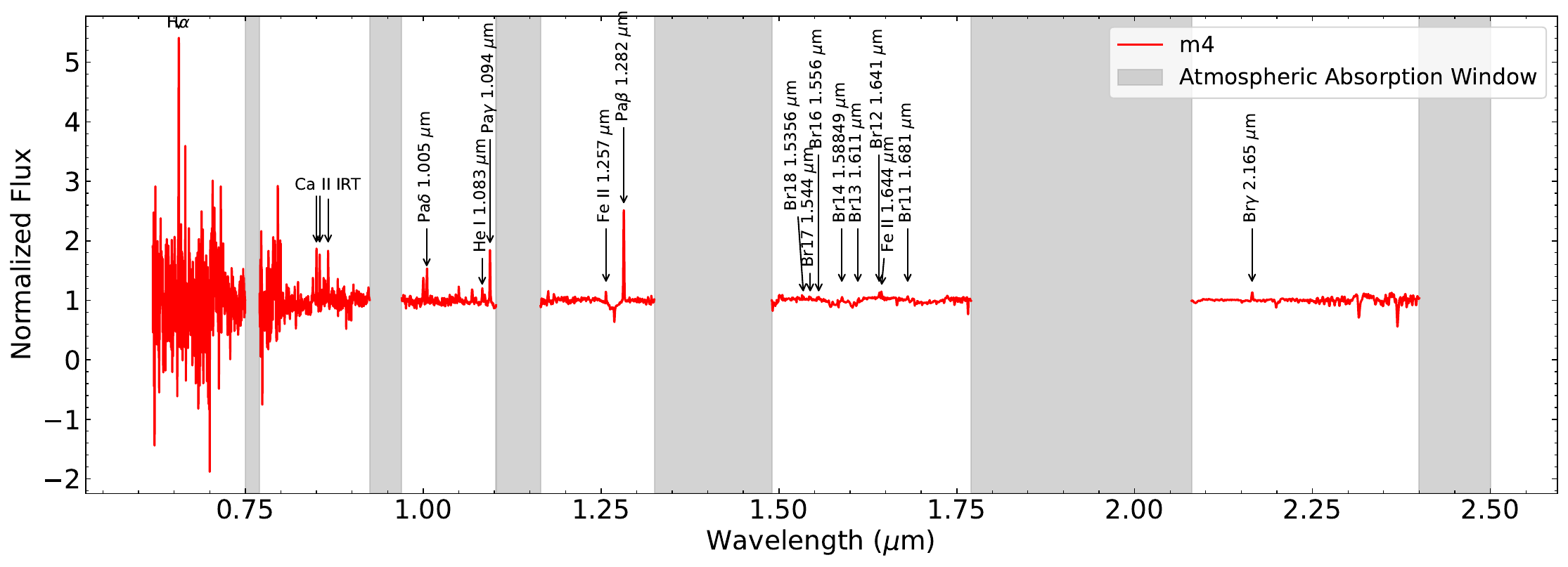}
    \caption{Upper Panel: Wavelength-calibrated normalized spectra of `m2 and m3', observed using HFOSC.
    Lower Panel: Wavelength-calibrated spectrum of m4, observed using TANSPEC. }
    \label{fig:hct_spectra}
\end{figure*}

The obtained wavelength-calibrated normalized spectra of `m2 and m3' are presented in the upper panel of Figure~\ref{fig:hct_spectra} with blue and green colors, respectively. The spectrum of `m1' was not used in our analysis as the signal-to-ratio was too weak.
We applied various spectral libraries and criteria reported in the literature for the spectral classification of probable massive stars (see also \citealt{1984ApJS...56..257J,1990PASP..102..379W,2020ApJ...894....5R}). The spectra of OB-type stars consist of several hydrogen, and helium lines along with some other atomic lines, such as C\,{\sc iii}, Mg\,{\sc ii}, O\,{\sc ii}, Si\,{\sc iii}, Si\,{\sc iv}. The strength of the He\,{\sc ii} line gets weaker for late O-type stars, and it is last visible in B0.5-type stars \citep{1990PASP..102..379W}. Since we could not find these lines, we determined that the spectral types of `m2 and m3' are later than B0.5. Then we applied the criteria reported by \citet{2020ApJ...894....5R} on the spectra of `m2 and m3' and searched for the presence of Balmer lines at 4104, 4340, 4860, and 6562 \r{A} and Si\,{\sc iv} line at 4089 \r{A}, that exist in the spectrum of `m2' whereas are absent in the spectrum of `m3'. The presence of these lines in the spectrum of `m2' implies that its spectral type is between B0 and B2, whereas the absence in the spectra of `m3' implies that its spectral type is later than B2. Then for `m2', we further compared its spectrum with the spectral library reported by \citet{1984ApJS...56..257J} and concluded its spectral type as B1.5. Following the criteria reported by \citet{2020ApJ...894....5R} for `m3', we find that the He\,{\sc i} line at 4471 \r{A} is stronger than the Mg\,{\sc ii} line at 4481 \r{A}, so we constrain its spectral type between B3-B5. Further comparing its spectrum with the spectral library reported by \citet{1984ApJS...56..257J}, we conclude its spectral type is B3.

Since we did this classification using low-resolution spectra, we expect an uncertainty of $\pm$1 in the classification of its subclass.

%%%%%%%%%%%%%%%%%%%%%%%%%%%%%%%%%%%%%%%%

\subsection{Spectroscopic Analysis of the MYSO `m4' in optical-NIR regime}

Ionized atomic emission lines provide insights into the excitation conditions of the YSOs and its environment. Key features also serve as tracers for various phenomena, such as accretion in the inner disc region (e.g., Br$\gamma$), the presence of outflows (e.g., H$_2$ and shocked [Fe {\sc ii}]), and circumstellar discs, indicated by CO bandhead emission and fluorescent Fe {\sc ii} lines \citep{2022MNRAS.517.1518J}.  The NIR spectrum of `m4' has been presented by \citet{2013MNRAS.430.1125C} using United
Kingdom Infra-Red Telescope (UKIRT) imager-spectrometer (UIST) instrument, but with a low spectral resolution ($\lambda/\Delta\lambda$) of $\sim$500. So we reobserved it through TANSPEC, serving a comparatively higher spectral resolution of $\sim$1500. We present the wavelength-calibrated normalized spectrum of `m4' in the lower panel of Figure~\ref{fig:hct_spectra} with red color. The aim is to confirm the nature of MYSO using optical-NIR emission features. The spectrum exhibits a diverse range of emission lines from atomic and molecular species. We observe a strong He {\sc i} emission line at 1.083 $\mu$m, which is likely due to a composite origin,  including contributions from an accretion shock, from the funnel flow, and from wind \citep{2003ApJ...599L..41E}. We observed a Br$\gamma$ emission line at 2.165 $\mu$m, which is also a crucial tracer of magnetospheric accretion \citep{2018ApJ...861..145C}. We also observe strong hydrogen recombination lines from the Brackett (Br) series at 1.5346, 1.544, 1.556, 1.58849, 1.611, 1.641, and 1.681 $\mu$m along with Paschen (Pa) series in emission (Pa$\beta$ at 1.282 $\mu$m, Pa$\delta$ at 1.005 $\mu$m, and Pa$\gamma$ at 1.094 $\mu$m), which are significant characteristic tracers of magnetospheric accretion \citep{2023JApA...44...50G}. We observed the Ca {\sc ii} IR triplet (IRT) emission lines at around 0.8500, 0.8545, 0.8664 $\mu$m, which are considered to form in the magnetospheric accretion process \citep{1998AJ....116..455M}. We observe the H$\alpha$ emission line at 0.6563 $\mu$m, which suggests a combination of hot spots, the accretion process, stellar rotation, and magnetic field topology, in addition to the accretion rate \citep{2014A&A...561A...2A}. We observe some metallic lines also, such as [Fe {\sc ii}] at 1.257 and 1.644 $\mu$m, which are usually found in the shock-excited gas and protostellar jets \citep{2013ApJ...778...71G,2016PASP..128g3001P} and supernova remnants \citep{2019AJ....157..123L}. The protostellar jets from young stars significantly influence star formation and disk evolution dynamics. They control the process of stellar accretion by eliminating the angular momentum generated by the accreting material in the disk and customizing the inner disk physics, which influences the evolution of proto-planetary systems. The shaded region depicts the atmospheric absorption window at Maunakea\footnote{\url{http://twiki.cis.rit.edu/twiki/bin/view/Main/MaunaKeaTo100kmAtmosphericTransmissions}}.
Due to several accretion tracer lines, we conclude that the MYSO `m4' is undergoing accretion processes. One important finding to highlight here is that we do not observe any P-Cygni profile, as reported by \citet{2013MNRAS.430.1125C}; though we have observed it with a higher resolution ($\sim$1500) TANSPEC instrument than UIST imager-spectrometer (resolution $\sim$500). Further investigation on `m4' is beyond the scope of this study.

\subsection{Stellar Clustering associated with the E71 Bubble}\label{sec:clustering}

\citet{2013yCat..35580053K} reported a stellar cluster within E71 at a distance of 1.715 kpc, with an extinction value ($E(B-V)$) of 1.02 mag, derived from isochrone fitting on 2MASS data.
% The nearest neighbor (NN) method is used to decipher this region's structure and stellar clustering.
We employed the nearest neighbor (NN) technique, as described by \citet{2005ApJ...632..397G}, on an NIR catalog compiled from 2MASS and UKIDSS data. Sources with \emph{J}-band magnitudes $\leq$13 were taken from 2MASS, while fainter sources (\emph{J} > 13) were adopted from UKIDSS. This technique was used to investigate the structure and stellar clustering within the region. We derived a stellar surface density map to decipher the stellar clustering associated with the E71 bubble. For this, we fluctuated the radial distance with a grid size of 6$\arcsec$ such that it encloses the twentieth nearest star. The stellar surface density $\sigma$ at [i, j] grid position is calculated by:

\begin{equation}
    \sigma(i,j) = \frac{N}{\pi r_N^2(i,j)}
\end{equation}

Here $r_N^2(i,j)$ denotes the projected radial distance of grid position ([i, j]) to the \emph{N$^{th}$} nearest star.

We overlaid the surface density contours on the lower right panel of Figure~\ref{fig:intro} with cyan color. The lowest level is 2$\sigma$ above the mean value (6.36 stars arcmin$^{-2}$), with a step size of 1$\sigma$ (1.14 stars arcmin$^{-2}$). A clear clustering of stars is visible inside the ring of dust and gas at $\alpha_{J2000}=05^h38^m43^s$.8 and $\delta_{J2000}=+30^{\circ}43\arcmin43\arcsec$. 
%Due to this elliptical morphology, 
We define the area of the cluster ($A_{cluster}$) through its \emph{convex hull}\footnote{\emph{Convex hull} is a polygon enclosing all points in a grouping with internal angles between two contiguous sides of less than 180$^{\circ}$.} (or Qhull) using the formula (\citealt{2006A&A...449..151S}, \citealt{2016AJ....151..126S}, \citealt{2020MNRAS.498.2309S}),

\begin{equation}
    \begin{split}
        A_{cluster}=\frac{A_{hull}}{1-\frac{n_{hull}}{n_{total}}}
    \end{split}, 
\end{equation}

Where $n_{hull}$ is the total number of objects on the perimeter of the hull, and $n_{total}$ is the total number of objects inside the hull. 
%The idea behind redefining the cluster area with a convex hull instead of a circular one is that stellar clusters generally have an elliptical morphology, and defining them with a circular area brings an overestimation \citep{2006A&A...449..151S,2016AJ....151..126S}. 
We have shown this cluster's \textit{convex hull} (blue polygon) in the lower right panel of Figure~\ref{fig:intro} and will refer to it as `Cl1' in our study. $A_{cluster}$ is measured as 18 arcmin$^2$ for Cl1. 
The radius of the Cl1 ($R_{cluster}$), which is the radius of the circle having an area equal to $A_{cluster}$, is then estimated as $=2\arcmin.4=1.26$ pc, considering the cluster is located at a distance 1.81 kpc (see Section~\ref{sec:membership}). 
% This radius is smaller than the reported radius of the E71 bubble ($=1\arcmin.65$, cf. Section \ref{sec:intro}, which corresponds to $\sim$1.58 pc for a distance 1.81 kpc). 
As can be observed from the lower-right panel of Figure~\ref{fig:intro}, the cluster Cl1 is located inside the bubble of gas and dust. $R_{circ}$ is defined as half of the farthest distance between two hull objects/stars, which comes out to be $2\arcmin.23=1.17$ pc.
The aspect ratio 
% ($\frac{R^2_{circ}}{R^2_{cluster}}$) 
($=R^2_{circ}/R^2_{cluster}$) of Cl1 is then calculated as 0.9, indicating more or less this cluster's circular morphology. The physical parameters for Cl1 are tabulated in Table~\ref{tab:mst}.

%%%%%%%%%%%%%%%%%%%%%%%%%%%%%%%%%%%%%%%%

\subsection{Membership Probability Analysis and Distance Estimation of the Bubble} \label{sec:membership}

Distance estimation is crucial for constraining various physical parameters of star clusters, such as their size, mass, and age. According to \citet{2019PASJ...71....6H}, E71 is located at a distance of 3.3 $\pm$ 0.4 kpc, with an angular radius of 1$\arcmin$.65 and a covering fraction of 0.54. 

To re-estimate the distance of the cluster associated with the E71 bubble, referred to as Cl1 (see Section~\ref{sec:clustering}), we utilized the most precise parallax measurements available to date from \textit{Gaia} DR3 \citep{2023gaia}. We considered stars with parallax errors $\leq 0.2$ mas within a $9\arcmin \times 9\arcmin$ region, which encompasses the entire bubble as well as the locations of m1', m2', m3', and m4'. Converting parallax measurements into distance estimates is non-trivial due to the need for positive distances and the non-linear nature of the transformation \citep{2018A&A...616A...9L}. The most effective method for addressing this is through a probabilistic approach. \citet{2021AJ....161..147B} proposed a robust method that uses parallax to generate the full likelihood distribution for distances. An added benefit of this approach is its lower uncertainty compared to kinematic distance estimates, especially for stars located within 5 kpc \citep{2022A&A...663A.133K}.

\begin{figure}[!ht]
    \centering
    \includegraphics[width=\linewidth]{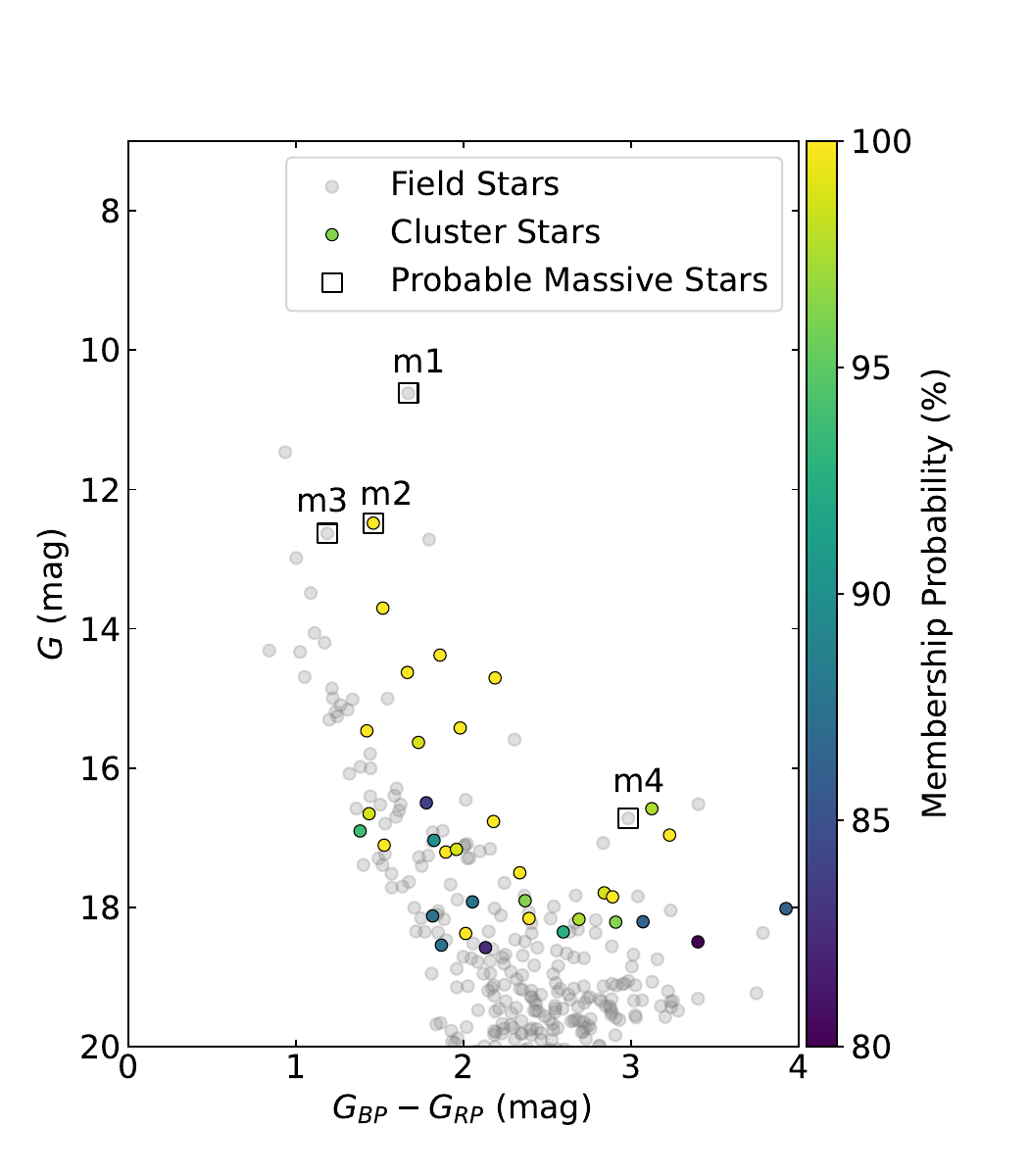}
    \caption{\emph{Gaia} DR3 \textit{G} vs. $(G_{BP}-G_{RP})$ CMD where the color bar represents the membership probabilities (in \%) computed by {\sc fastmp} in a $9\arcmin\times9\arcmin$ region. The locations of `m1, m2, m3, and m4' are also marked.}
    \label{fig:prob_dist}
\end{figure}

We estimated the membership probabilities of stars using the supervised algorithm {\sc fastmp} (fast Membership Probabilities), which does not require the prior selection of field stars \citep{2023MNRAS.526.4107P}. This represents an advantage over several earlier methods that necessitated such a selection \citep{1998A&AS..133..387B, 2004A&A...426..819B}. Figure~\ref{fig:prob_dist} displays the \emph{Gaia} DR3 \textit{G} vs. $(G_{BP}-G_{RP})$ CMD, where the color bar denotes the membership probabilities assigned by {\sc fastmp}. Stars with membership probabilities $\geq 80\%$ were identified as cluster members and are marked accordingly on the CMD.
It is worth noting that the identified cluster members are predominantly bright sources. This is a direct consequence of the stringent parallax error criterion we adopted—specifically, limiting parallax uncertainties to $\leq 0.2$ mas.
We determined the mean distance to the cluster members in Cl1 as $d = 1.81 \pm 0.15$ kpc. The cluster center identified by {\sc fastmp} is located at $\alpha_{J2000} = 05^h38^m51^{s}.1$ and $\delta_{J2000} = +30^{\circ}43\arcmin49\arcsec.8$ (marked as a blue $\times$ in Figure~\ref{fig:environment}(b)), which is consistent with the cluster center derived from NIR data (see Section~\ref{sec:clustering}).
Interestingly, our analysis classifies star `m2' as a cluster member, while `m1', `m3', and `m4' are not (cf. Figure~\ref{fig:prob_dist}). `m1' and `m3' may thus belong to the field population. `m4', being a deeply embedded MYSO, likely suffers from unreliable parallax or photometric measurements in optical surveys. Given its confirmed membership, `m2'—the only massive star among the identified members—could serve as the ionizing source of the E71 bubble.

\begin{figure}[!ht]
    \centering
    \includegraphics[width=0.88\linewidth]{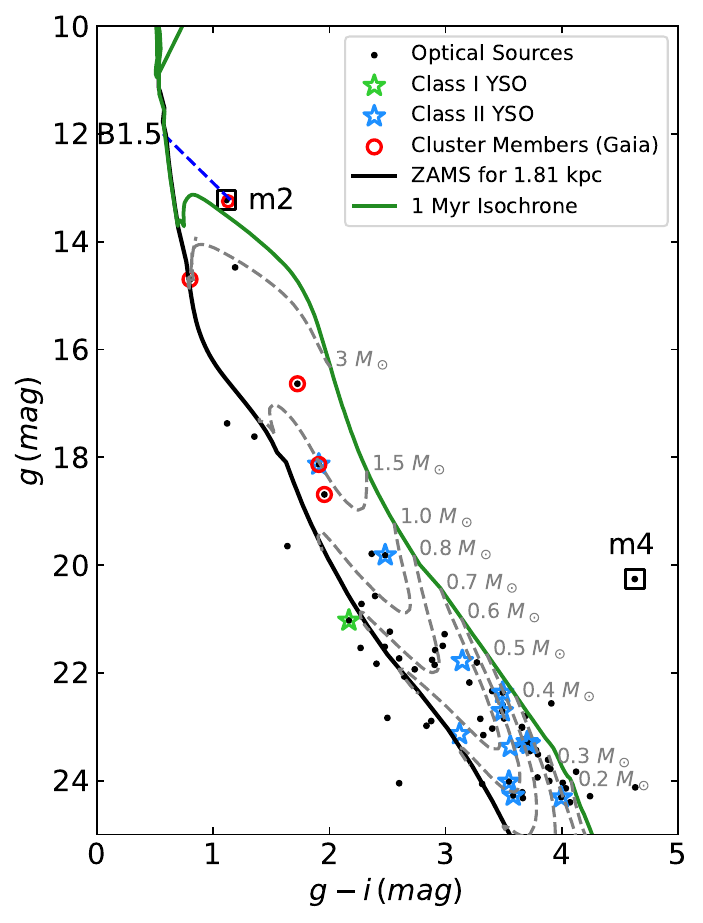}
     \caption{$g$ versus $(g-i)$ CMD for statistically cleaned stars within Cl1. The black and green solid curves represent the ZAMS and 1 Myr isochrones by \citet{2019MNRAS.485.5666P}, corrected for the distance 1.81 kpc and $A_V$ =2.63 mag. The blue dashed line represents the reddening vector. The evolutionary tracks of different masses are also shown with grey color. The cluster members (red circles) and `m2' and `m4’ are also marked.}
    \label{fig:dot_cmd}
\end{figure}

To validate this distance estimate, we analyzed the statistically cleaned $g$ versus $(g-i)$ CMD shown in Figure~\ref{fig:dot_cmd}, constructed using deep optical photometric observations from the 3.6-m DOT. The CMD has been statistically decontaminated by subtracting the contribution of field stars, estimated from the CMD of a nearby reference field (see \citealt{2017MNRAS.467.2943S} for methodological details). The CMD displays the distribution of the identified cluster members, along with the intrinsic zero-age main sequence (ZAMS; solid black curve) from \citet{2019MNRAS.485.5666P}, corrected for a distance of 1.81 kpc and a visual extinction of $A_V = 2.63$ mag. The adopted extinction value was derived from the Bayestar19 3D dust map \citep{2019ApJ...887...93G, 2019ApJ...877..116W}. The ZAMS provides a good fit to the observed CMD, thus reinforcing the distance estimate of 1.81 kpc for the Cl1 cluster (for details of CMD and isochrone fitting, see \citealt{1994ApJS...90...31P}).
The positions of stars ‘m2’ and ‘m4’ on the CMD further support their classifications as a B1.5-type star and a very young, massive stellar object, respectively.

%%%%%%%%%%%%%%%%%%%%%%%%%%%%%%%%%%%%%%%%

\subsection{Initial Mass Function (IMF)}\label{sec:mass_func}

The Initial Mass Function (IMF) is a fundamental statistical tool used to understand star formation within a given volume of space. It describes the distribution of stellar masses formed during a single star formation event. Mathematically, it is expressed as a power law $N(log\,m) \propto m^{\Gamma}$ with its slope defined as
\begin{equation}
    \Gamma = \frac{d\,log\,\phi}{d\,log\,m}.
\end{equation}

The notation $\phi = N(\log m)$ represents the total number of stars per unit interval in logarithmic mass. We used the $g$ versus $(g-i)$ CMD generated through the deep optical data from IMAGER mounted at 3.6-m DOT (Figure~\ref{fig:dot_cmd}) corrected for the field star contamination and incompleteness of the data (see Section~\ref{sec:completeness}). We adhered to the procedures described in the works of \citet{2020ApJ...891...81P} and \citet{2017MNRAS.467.2943S}. Briefly, we constructed CMDs of $g$ versus $(g - i)$ for both the cluster and a nearby reference field. We then divided both CMDs into uniform grids with bin sizes of $\Delta g = 1$ mag and $\Delta(g - i) = 0.5$ mag. Within each bin, we calculated the number of stars in both the cluster and field CMDs and estimated the number of likely cluster members by subtracting the field counts from the cluster counts. To statistically clean the cluster CMD, we removed a corresponding number of stars—those closest in position within the CMD—from the cluster field, effectively minimizing the contribution of unrelated foreground and background stars.

\begin{figure}[!ht]
    \centering
    \includegraphics[width=\linewidth]{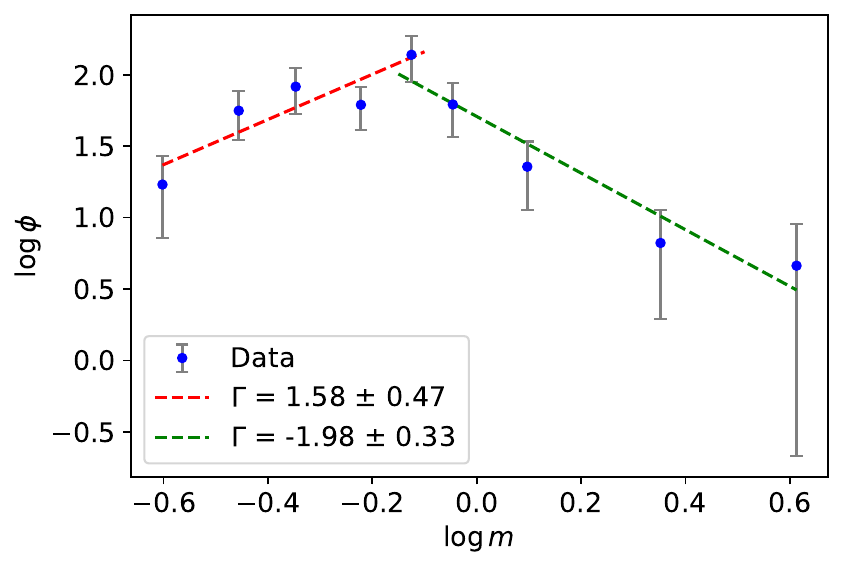}
    \caption{A plot of the MF distribution for the stars within Cl1 identified using deep optical data from IMAGER mounted at 3.6-m DOT. Here, $\phi$ marks $N(\log m)$, the error bars mark $\pm\sqrt N$ errors. The red and green dashed line shows the least squares fit to the MF distribution.}
    \label{fig:mf}
\end{figure}

It is reported that the higher-mass stars predominantly follow the classical Salpeter Mass Function (MF; \citealt{1955ApJ...121..161S}). In contrast, the IMF becomes less well-constrained at lower masses, where it tends to flatten below 1 M$_\odot$ and shows a relative deficit of stars at the lowest mass end \citep{2002Sci...295...82K, 2003PASP..115..763C, 2015arXiv151101118L, 2016ApJ...827...52L}. Our analysis yields a distinct change in the slope of MF ($\Gamma$) at $\sim$0.7 M$_\odot$ (Figure~\ref{fig:mf}). Such a break in the MF slope has also been reported in previously observed clusters \citep[e.g.,][]{ Sharma_200710.1111/j.1365-2966.2007.12156.x, Jose_2008}.  We defined the bins in $log\,m$ in such a way that their size in log scale should be uniform and have at least 2 stars in them. By using least-squares fitting, we found that $\Gamma$ is $-1.98 \pm 0.33$ within the mass range $\rm \sim0.7 < M/M_\odot < 5.2$, while it is $+1.58 \pm 0.47$ in the mass range $\rm \sim0.2 < M/M_\odot < 0.7$.

%%%%%%%%%%%%%%%%%%%%%%%%%%%%%%%%%%%%%%%%

\subsection{Extraction of the YSOs Embedded in the Bubble}\label{sec:mst}

\begin{figure*}[!ht]
\begin{minipage}{0.6\textwidth}
\includegraphics[width=\textwidth]{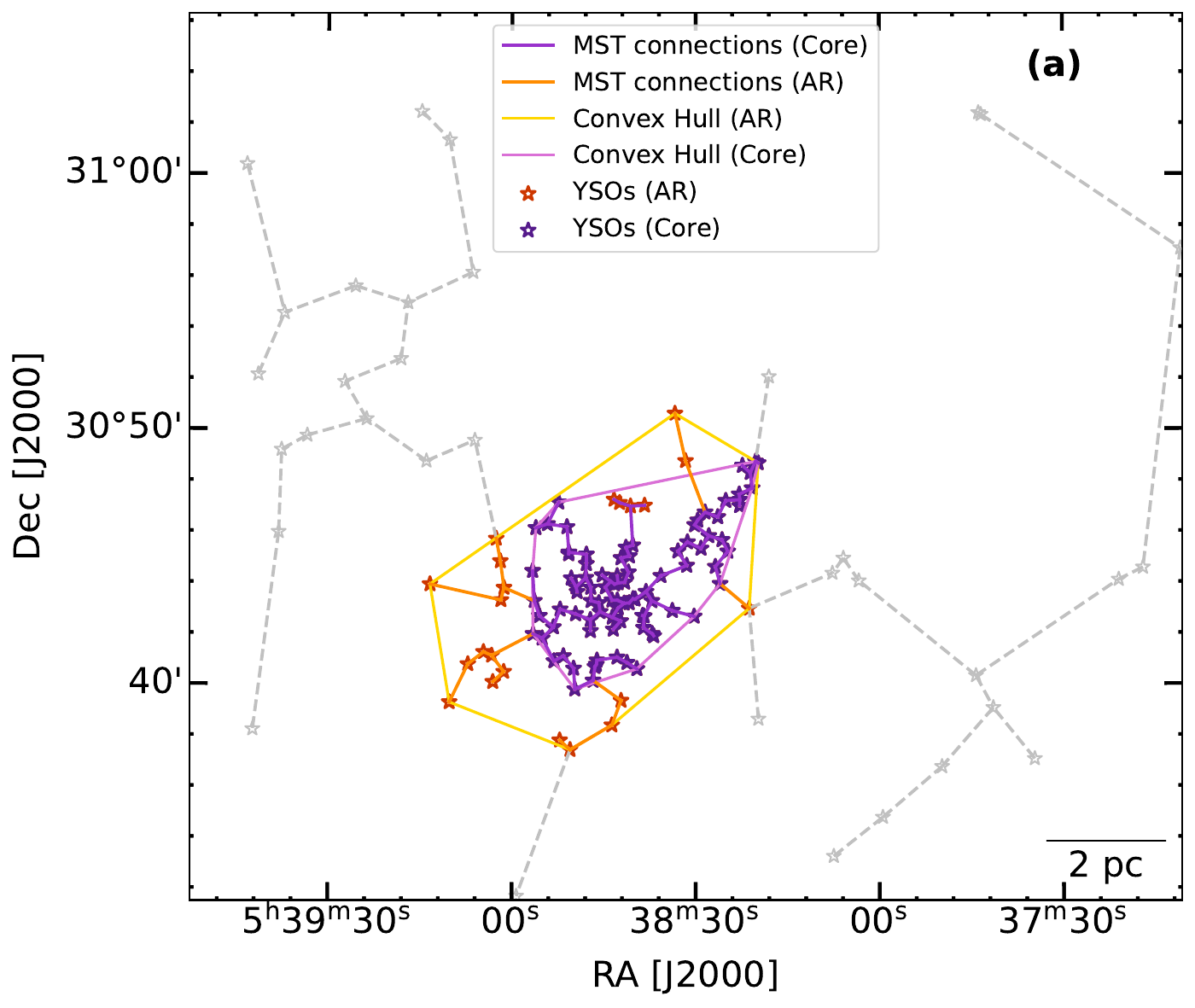}
\end{minipage}
\hspace{0.01\textwidth}
\begin{minipage}{0.35\textwidth}
\includegraphics[width=\textwidth]{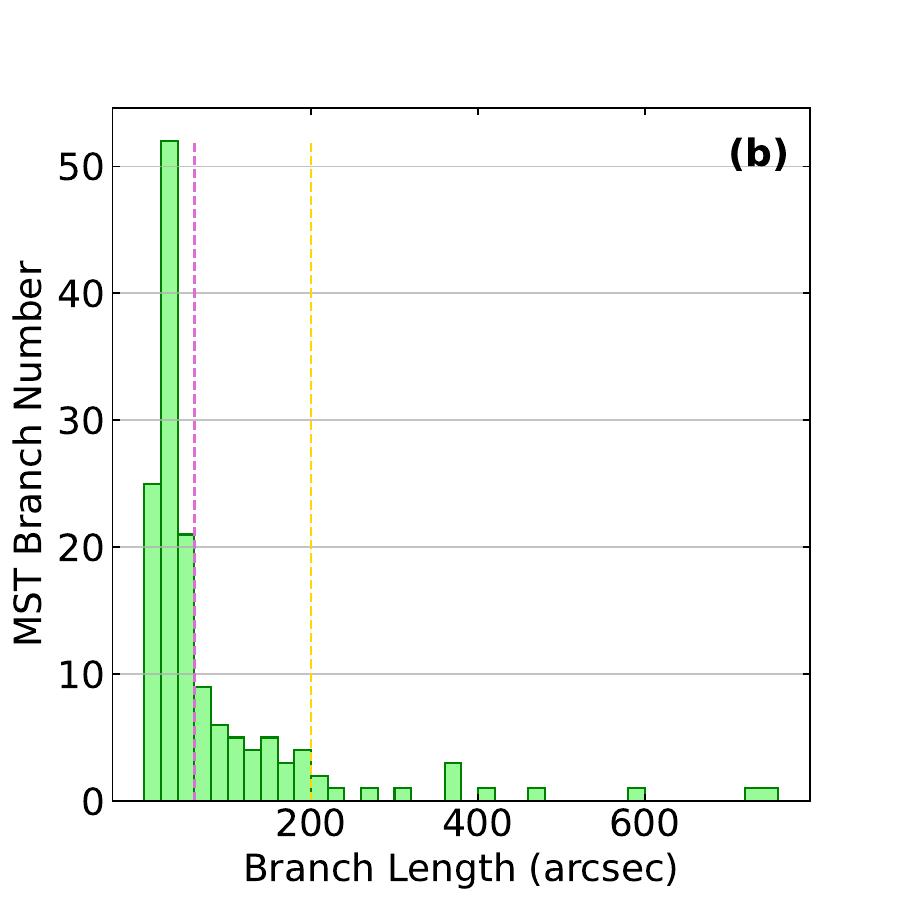}
\includegraphics[width=\textwidth]{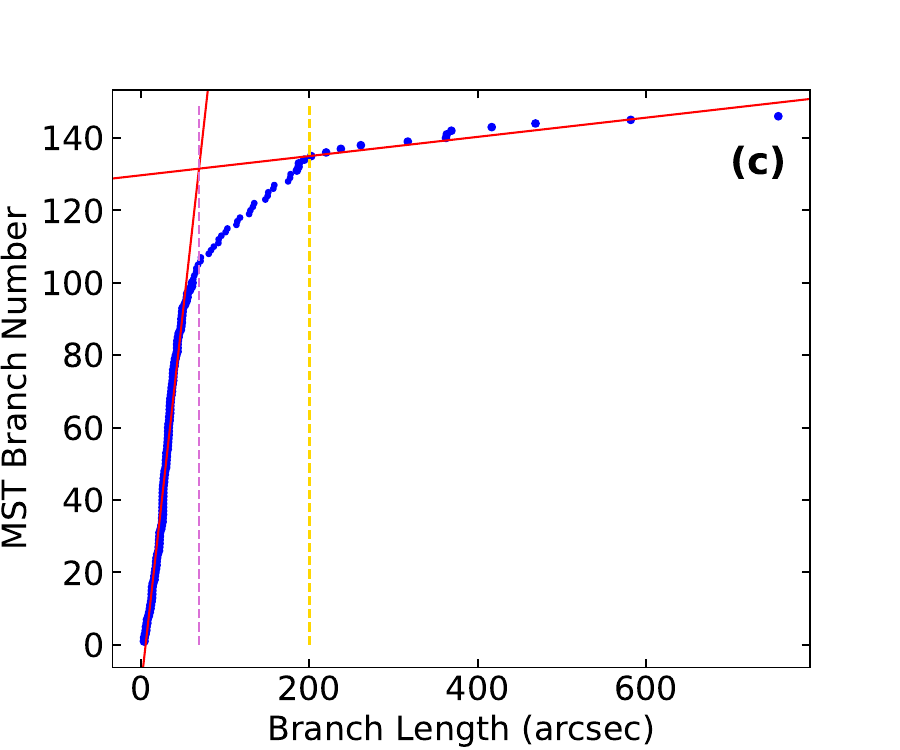}
\end{minipage}
\caption{(a) MST for the classified YSOs. MST connections are shown with grey color along with the isolated AR and core enclosed by the yellow-colored and magenta-colored \emph{Convex hulls}, respectively. The MST connections inside them are orange and purple, respectively. 
(b) and (c): Histogram and CS plot of the MST branch lengths, respectively, plotted to estimate the $l_{critical}$. The magenta and yellow vertical lines in both panels represent the $l_{critical}$ for extracting the core and AR, respectively.}
\label{fig:mst}
\end{figure*}

Since the E71 bubble shows signatures of recent star formation, a young stellar population, i.e., YSOs, should be associated with it. Since YSOs have a circumstellar disk around them, we identified them based on their excess IR emission (refer to Appendix~\ref{app:yso_identification} for further details). We uncovered that 8 Class $\textsc{i}$ and 139 Class $\textsc{ii}$ YSOs are present in the $35\arcmin\times35\arcmin$ area of our target region. We have shown this population in the lower left panel of Figure~\ref{fig:intro} with green and blue asterisks, respectively. We wanted to extract the core (bound groups of YSOs having similar star formation history) and active region (AR; a region with active star formation). To do so, we have used an empirical technique called \emph{`Minimal Spanning Tree'} (MST; \citealt{2009ApJS..184...18G}). This technique is used to separate out the groupings while preserving the underlying geometry. The extracted MST has been shown in Figure~\ref{fig:mst}(a). To extract the AR and core, we first plotted the histogram of the MST branch lengths (i.e., separation between YSOs, shown in Figure~\ref{fig:mst}(b)), which displays a peak at smaller lengths with a longer tail for larger MST branch lengths. This distribution points towards substantial subregion(s) above a fairly uniform and elevated surface density. By setting a threshold for MST branch lengths, we can identify sources that are closer than this limit, allowing us to isolate populations contributing to local surface density enhancements. To attain this threshold, we fitted two lines in the shallow and steep data points of the cumulative sum (CS) plot of the MST branch lengths (see Figure~\ref{fig:mst}(c)) and estimated the point of intersection of these two lines as the critical MST branch length ($l_{critical}$) for the core. We have plotted a magenta vertical line at this point. In like manner, we estimated the $l_{critical}$ for the AR by choosing a point in the cumulative sum plot where the less-sloped line exhibits a gap in the distribution of the MST branch lengths. $l_{critical}$ is measured to be $\sim 70\arcsec$ for core and $\sim 200\arcsec$ for AR. 
%This should be kept in mind that $l_{critical}$ is not a fixed value and may change slightly as the data statistics vary. 
After separating out these YSOs, we enclosed them with their respective convex hulls (shown with magenta color for core and yellow color for AR) in Figure~\ref{fig:mst}(a). 
%The main conclusion from this analysis is the observation of a significant level of clustering in our extracted AR/core. 
We then estimated the physical parameters of the identified core, AR, corona (i.e., the region which is outside the core but is enclosed by the AR), and is mentioned in Table~\ref{tab:mst} \citep[see][for methodologies]{2016AJ....151..126S}.

%The calculated aspect ratio for the core and the AR (1.84 and 1.59, respectively) hints towards their elongated morphology.
% The $R_{cluster}$ and $R_{circ}$ for the core are 2.12 and 2.87 pc, respectively; whereas for the AR, they are measured to be 2.98 and 3.75 pc, respectively. The aspect ratio of the core and AR comes out to be 1.84 and 1.59, respectively, which hints towards the elongated morphology of the cluster.

%We used the $Q$ parameter, defined as the ratio of the normalized mean MST branch ($\bar{l}_{MST}$) to the normalized mean separation ($\bar{s}$) between objects \citep{2006A&A...449..151S,2014MNRAS.439.3719C,Ascenso2018}, to evaluate the strength of hierarchical versus radial distributions of YSOs in the extracted groupings.

%\begin{equation}
%    Q=\frac{\bar{l}_{MST}}{\bar{s}}
%\end{equation}

%Here, mean MST branch (${l}_{MST}$) is normalized by a factor $\sqrt{A_{cluster}/n_{total}}$ whereas the mean separation (${s}$) is normalized by $R_{cluster}$. 

We also calculated the star formation efficiency (SFE; $\epsilon$) to explore the efficiency of the formation of new stars from the available molecular gas in Cl1, AR, and the core. It defines how effectively molecular gas mass is converted into stars,

\begin{equation}
    \epsilon = \frac{M_{Star}}{M_{Cloud}+M_{Star}}
\end{equation}

Here, $M_{Star}$ denotes the mass of the young stellar population, i.e., YSOs. Since most of the YSOs lie in the faint magnitude limits (see Figure \ref{fig:dot_cmd}), we considered a typical value of $M_{Star}$ as 0.5 $M_\odot$ \citep{2014MNRAS.439.3719C,2016AJ....151..126S}. $M_{Cloud}$ denotes the mass of the molecular cloud and is given as \citep{2017ApJ...845...34D}:

\begin{equation}\label{eq:mass}
    M =  \mu_{H_2}\,m_H\,a_{pixel}\,\Sigma N(H_2)
\end{equation}

Here, $\mu_{H_2}$ denotes the mean molecular weight per hydrogen molecule (i.e., 2.8), 
% $m_H$ denotes the mass of the hydrogen atom,
$a_{pixel}$ denotes the area subtended by 1 pixel, and $N(H_2)$ denotes the total column density of that particular region (we used \textit{Herschel} column density values for this purpose).

Another key parameter to infer the primitive structure of the star-forming regions is the Jeans fragmentation/Jeans length ($\lambda_J$), which is defined as the minimum radius of a homogeneous isothermal molecular cloud of mean temperature $T$ and density $\rho_0$ to undergo a gravitational collapse. Mathematically \citep{2014MNRAS.439.3719C}, 

\begin{equation}
    \lambda_J = \left(\frac{15 k T}{4 \pi G m_H \rho_0}\right)^{1/2}, 
\end{equation}

Here, $m_H$ refers to the atomic mass of the hydrogen, and $\rho_0$ is the mean density expressed as,

\begin{equation}
    \rho_0 = \frac{3 M_T}{4 \pi R_H^3}. 
\end{equation}

Here, $M_T$ represents the total mass of the cluster (molecular cloud and stars), and $R_H$ is the radius of the cluster ($=R_{cluster}$).

All of the above parameters for Cl1/core/AR/corona are listed in the Table~\ref{tab:mst}.

\begin{table*}[!ht]
    \footnotesize
    \centering
    \caption{Various parameters extracted for Cl1, AR, Core, and Corona (the region which is outside the core but is enclosed by the AR). 
    %\textbf{The values of the nearest neighbour distance and the mean branch lengths have been directly measured from the projected separation between YSOs.}
    %
    %It is to be noted that we have not reported the mean branch length and the Q parameter for Cl1 as they are the properties extracted for MST.
    }
    \begin{tabular}{c c c c c}
    \hline
    Parameters & Cl1 & Core & AR & Corona\\
    \hline
    YSOs enclosed within the convex hull & 30 & 91 & 113 & 22\\
    Fraction of Class $\textsc{i}$ YSOs & 0.03 & 0.04 & 0.05 & 0.09\\
    $R_{cluster}$ (pc) & 1.26 & 2.12 & 2.98 & -\\
    $R_{circ}$ (pc) & 1.17 & 2.87 & 3.75 & -\\
    Aspect Ratio & 0.86 & 1.84 & 1.59 & -\\
   % Nearest Neighbor Distance (pc) & 0.01 & 0.03 & 0.03 & 0.14\\
    Mean Branch Length/Separation between YSOs (pc) & - & 0.26 & 0.33 & 0.62\\
  %  Q Parameter & - & 0.68 & $\gtrapprox$0.8 & -\\
    Mass ($\times 10^2 M_\odot$) & 1.53 & 7.39 & 14.20 & 12.67\\
    $\lambda_J$ (pc) & 1.35 & 1.35 & 1.61 & -\\
    SFE (\%) & 12.82 & 8.46 & 5.65 & 2.38\\
    \hline\\
    \end{tabular}
   
    \label{tab:mst}
\end{table*}

%%%%%%%%%%%%%%%%%%%%%%%%%%%%%%%%%%%%%%%%

\subsection{Physical Environment around E71 Bubble}

We used ancillary data sets ranging from optical to radio to better comprehend the star formation activities (such as the distribution of YSOs, dust, gas, ionized gas, etc.) in the target region. The emission at longer wavelengths is primarily thermal emission from cold, dense gas. In contrast, the emission at shorter wavelengths is predominantly due to photospheric emission from the stellar sources.

\begin{figure*}[!ht]
    \centering
    \includegraphics[width=0.49\textwidth]{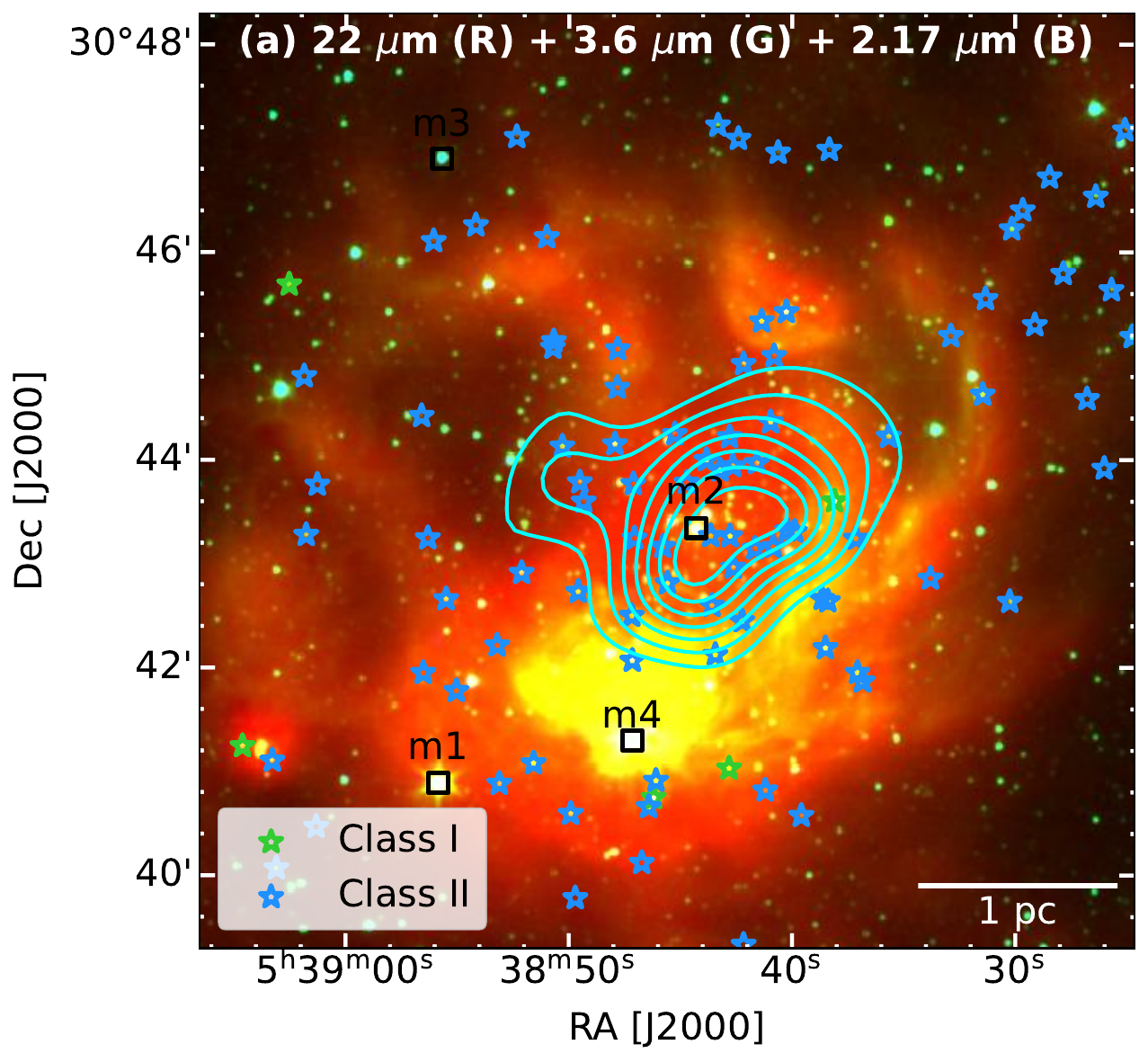}
    \includegraphics[width=0.49\textwidth]{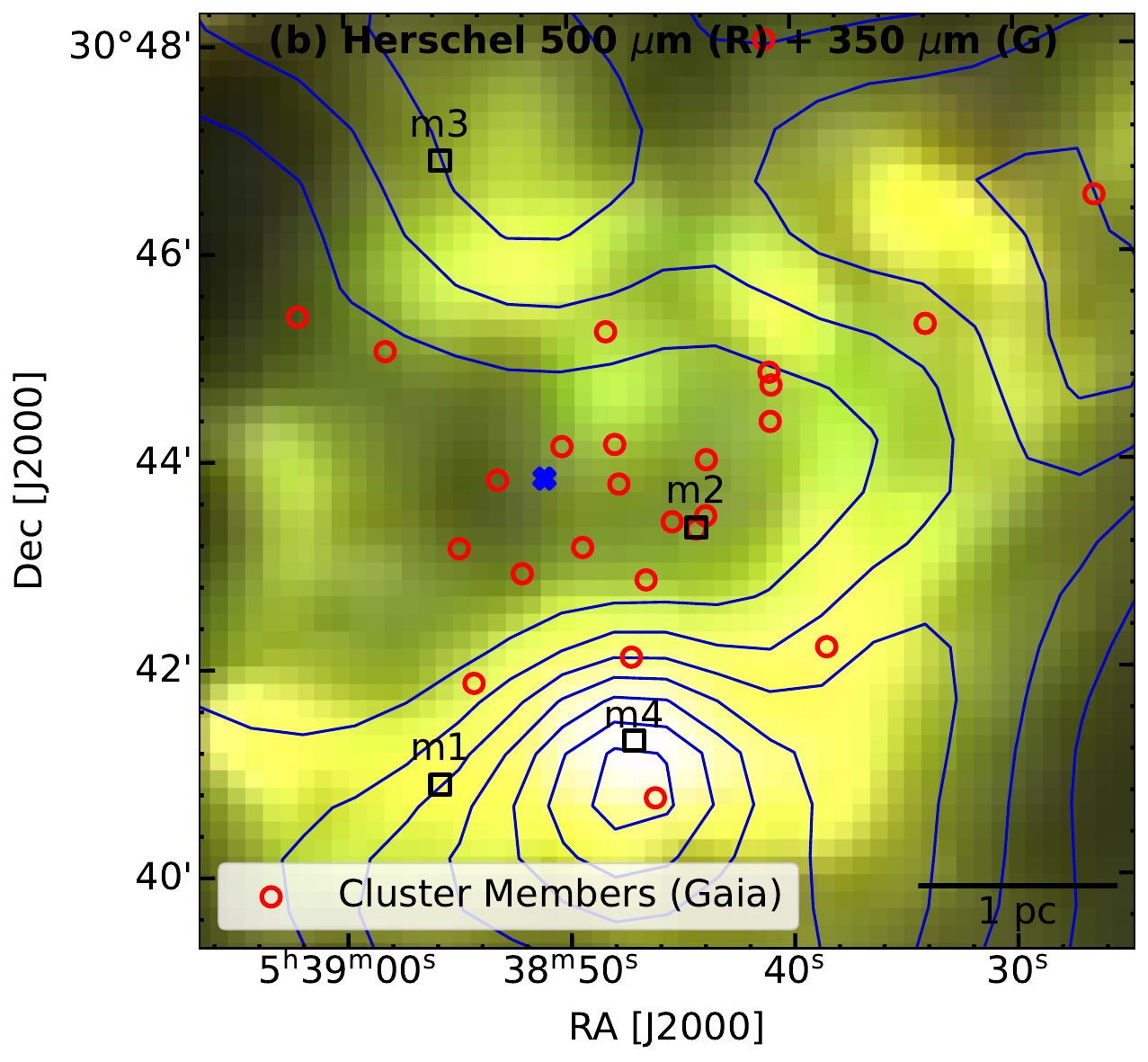}
    \includegraphics[width=0.49\textwidth]{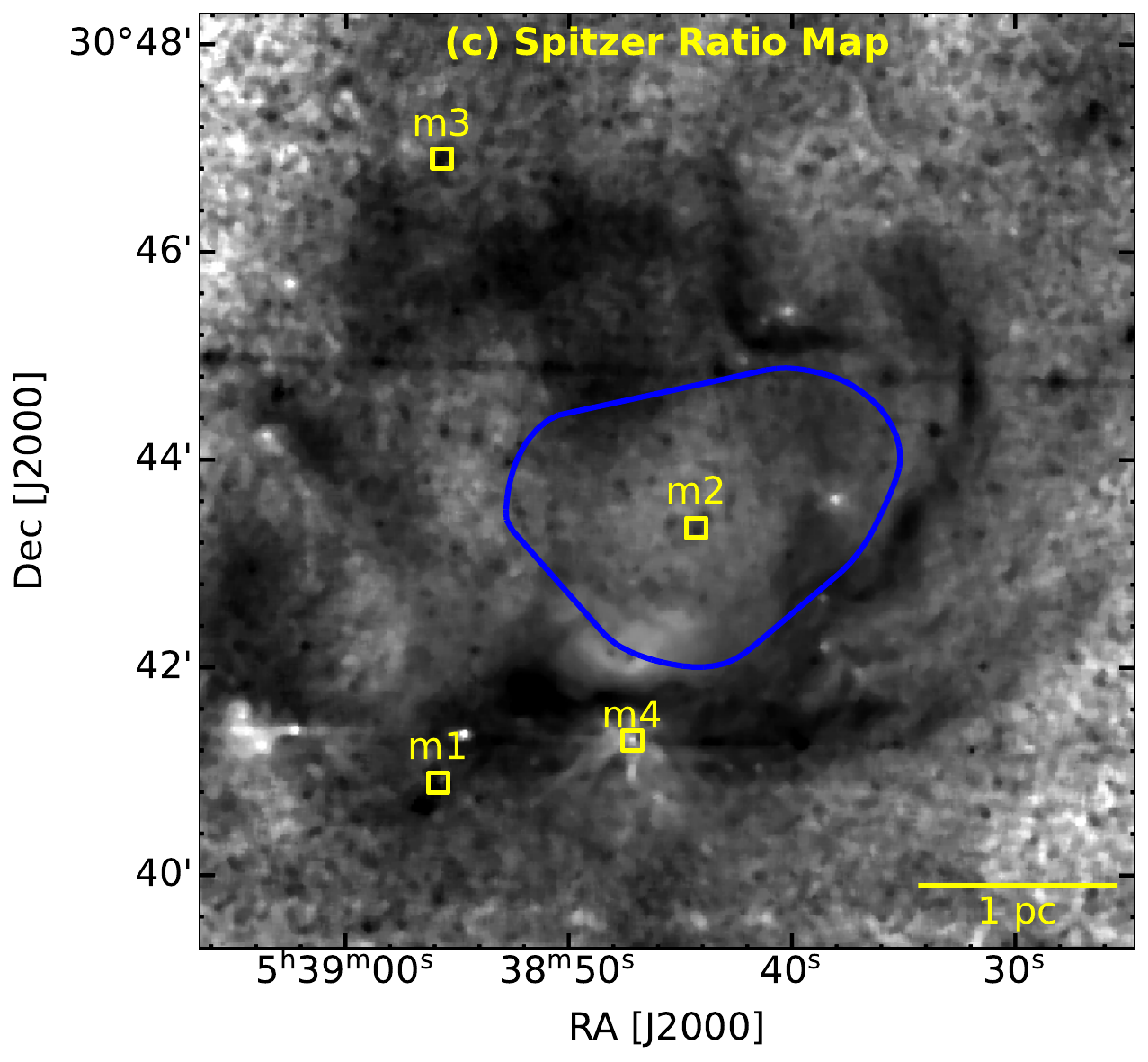}
    \includegraphics[width=0.49\textwidth]{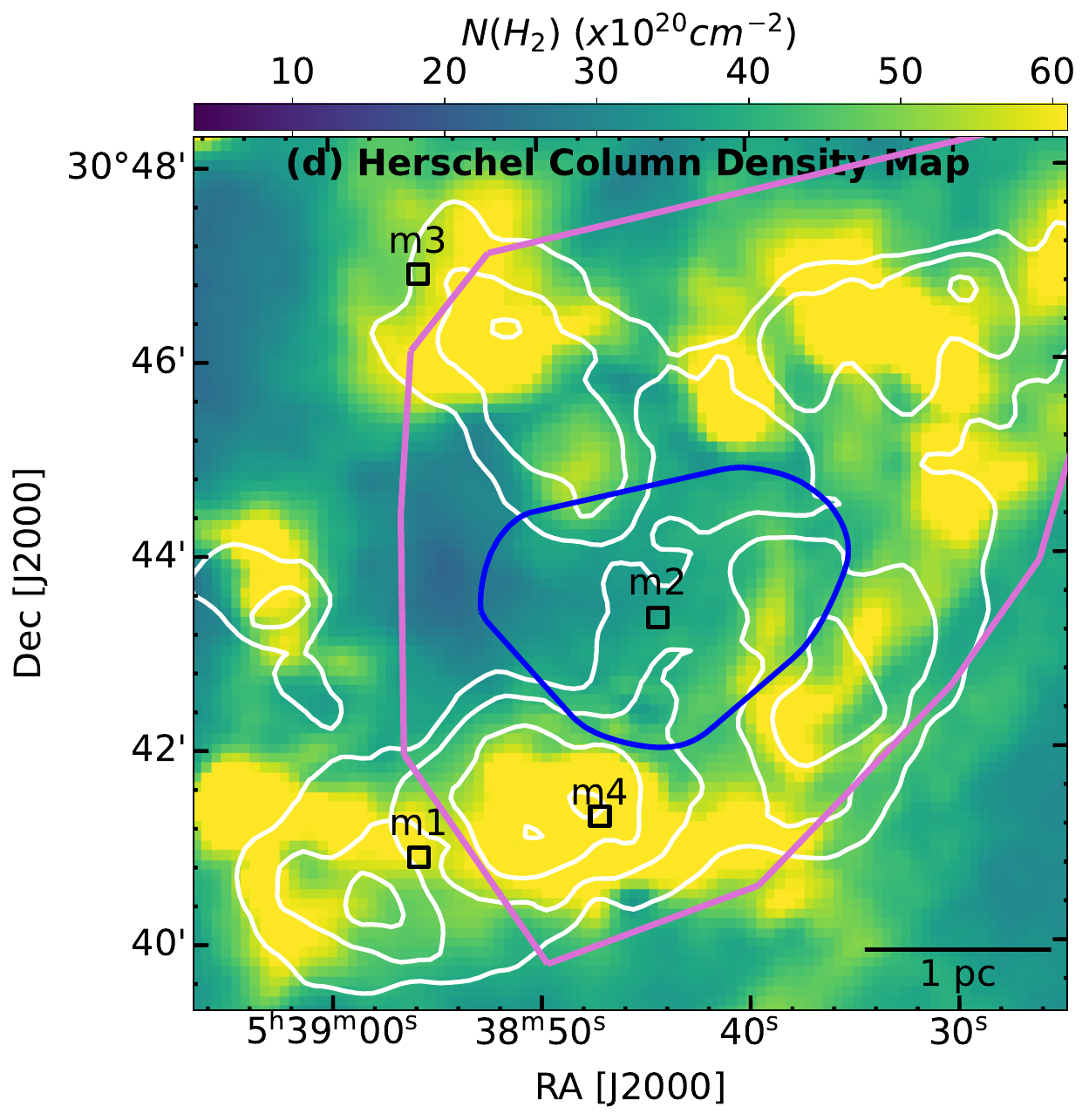}
    \caption{(a) Color-composite image generated using the WISE 22 $\mu$m, \textit{Spitzer} 3.6 $\mu$m, and 2MASS 2.17 $\mu$m (K-band) emission (red, green, and blue, respectively) overlaid with the isodensity contours (cyan) and the locations Class $\textsc{i}$ and Class $\textsc{ii}$ YSOs (green and blue asterisks), respectively. The lowest level for the isodensity contours is 6.36 stars arcmin$^{-2}$ with a step size of 1.14 stars arcmin$^{-2}$. 
    (b) Color-composite image generated using the \textit{Herschel} 500 $\mu$m, \textit{Herschel} 350 $\mu$m emission (red and green, respectively) overlaid with the locations of cluster members (red circles) and cluster center (blue '$\times$') identified using {\sc fastmp} algorithm. This map is also overlaid with the blue contours, representing $^{12}$CO integrated-intensity in the velocity ranges [$-20$, $-14$] \kms\, (c.f. Section~\ref{sec:ChannelMap}) where the lowest contours represent the emission above 5$\sigma$ value ($\sigma$ being the rms noise). The extent of these contours can be seen in the upper right panel of Figure~\ref{fig:intro}. (c) \textit{Spitzer} ratio map (4.5 $\mu$m/3.6 $\mu$m emission), smoothed using median smoothing of 5 pixels. 
    (d) \textit{Herschel} column density map overlaid with the extinction ($A_V$) contours (the lowest contour level is 8 mag with a step size of 0.85 mag) and the \textit{convex hull} of the core (magenta polygon; Section~\ref{sec:mst}). Panels (c) and (d) are overlaid with a blue polygon representing Cl1 (see Section~\ref{sec:clustering}).}
    \label{fig:environment}
\end{figure*}

Figure~\ref{fig:environment}(a) depicts the color-composite image of the WISE 22 $\mu$m, \textit{Spitzer} 3.6 $\mu$m, and 2MASS K (2.17 $\mu$m) band images overlaid with the surface density contours (cyan color; see Section~\ref{sec:clustering}) and identified YSOs (see Section~\ref{app:yso_identification}). The WISE 22 $\mu$m traces the distribution of warm dust emission, and the feedback from the massive star(s) might be one of the reasons for that. Whereas, the \textit{Spitzer} 3.6 $\mu$m image traces the PAH bands at 3.3 $\mu$m formed at PDRs. These PDRs might be created due to strong UV radiation from the massive star(s). The surface density contours seem entirely inside the partial ring (or bubble) of gas and dust. `m2' is found to be located near the peak of the stellar clustering. The spatial distribution of YSOs suggests that star formation activity is ongoing within the target bubble.

Figure~\ref{fig:environment}(b) depicts the color-composite image generated using the \textit{Herschel} 500 $\mu$m and 350 $\mu$m band images (red and green, respectively) and overlaid with the location of identified cluster members with membership probabilities $\geq 80\%$ (Section~\ref{sec:membership}). The \textit{Herschel} far-infrared (FIR) images reveal a partial ring or a shell-like structure of the E71 bubble
% that has also been marked with a red arc, 
with peak intensity coinciding with the location of MYSO `m4'.

Figure~\ref{fig:environment}(c) demonstrates the \textit{Spitzer} ratio map (4.5 $\mu$m/3.6 $\mu$m emission) smoothed using the median smoothing of five pixels. Since the \textit{Spitzer} 3.6 $\mu$m and 4.5 $\mu$m bands have similar point spread functions, they can be efficiently divided to omit the point sources and continuum emission (cf. \citealt{2017ApJ...834...22D}). This ratio map depicts some bright and dark regions; the darker regions point out the 4.5 $\mu$m emission due to prominent Br-$\alpha$ emission at 4.05 $\mu$m and a molecular hydrogen line emission ($\nu = 0-0 S(9)$) at 4.693 $\mu$m. In contrast, the brighter region points out the 3.6 $\mu$m emission due to PAH emission at 3.3 $\mu$m, which creates a PDR. This PDR may have resulted from the strong UV radiation emitted by the massive star(s) interacting with the surrounding molecular cloud. When we look carefully at the \textit{Spitzer} ratio map, we can observe a prominent distribution of PDRs, whose morphology is more or less analogous to a `lotus flower'. `m2' and the stellar clustering are located in the innermost cavity of this distribution, whereas `m4' is located just behind the PDRs.

We also generated the extinction ($A_V$) map using the $(H-K)$ colors of main-sequence (MS) stellar sources (excluding YSOs), employing the nearest-neighbor (NN) method \citep[cf.][]{2005ApJ...632..397G,2009ApJS..184...18G}. Since both the stellar surface density map and the $A_V$ map were derived from the same photometric catalog, they share comparable depth and sensitivity, allowing for a direct comparison between stellar distribution and the associated gas and dust content. A detailed description of the extinction map generation is available in \citet{2023ApJ...953..145V}.

Figure~\ref{fig:environment}(d) shows the \textit{Herschel} column density map overlaid with the extinction contours, enabling a clearer view of embedded structures. The extinction contours were generated starting from a base level of $A_V = 8$ mag, with a step size of 0.85 mag. Both the column density and $A_V$ maps reveal a partial ring-like (or arc-like) structure of gas and dust encompassing the stellar cluster. The MYSO ‘m4’ is notably located along this arc at a position of elevated column density.

Additionally, this map is overlaid with magenta and blue polygons representing the \textit{convex hulls} of the YSO core, identified through MST analysis, and the stellar cluster Cl1, derived from the NIR catalog (see Sections~\ref{sec:mst} and \ref{sec:clustering}, respectively). The extinction and stellar density maps clearly indicate that Cl1 is embedded within a region of elevated extinction, reflecting the influence of surrounding gas and dust. Furthermore, the spatial coincidence between the \textit{convex hull} of the YSO core and areas of higher column density suggests that a new generation of stars is actively forming within these denser environments.

%%%%%%%%%%%%%%%%%%%%%%%%%%%%%%%%%%%%%%%%

\subsection{Distribution of the Ionized Gas}

\begin{figure*}
    \centering
    \includegraphics[width=0.49\linewidth]{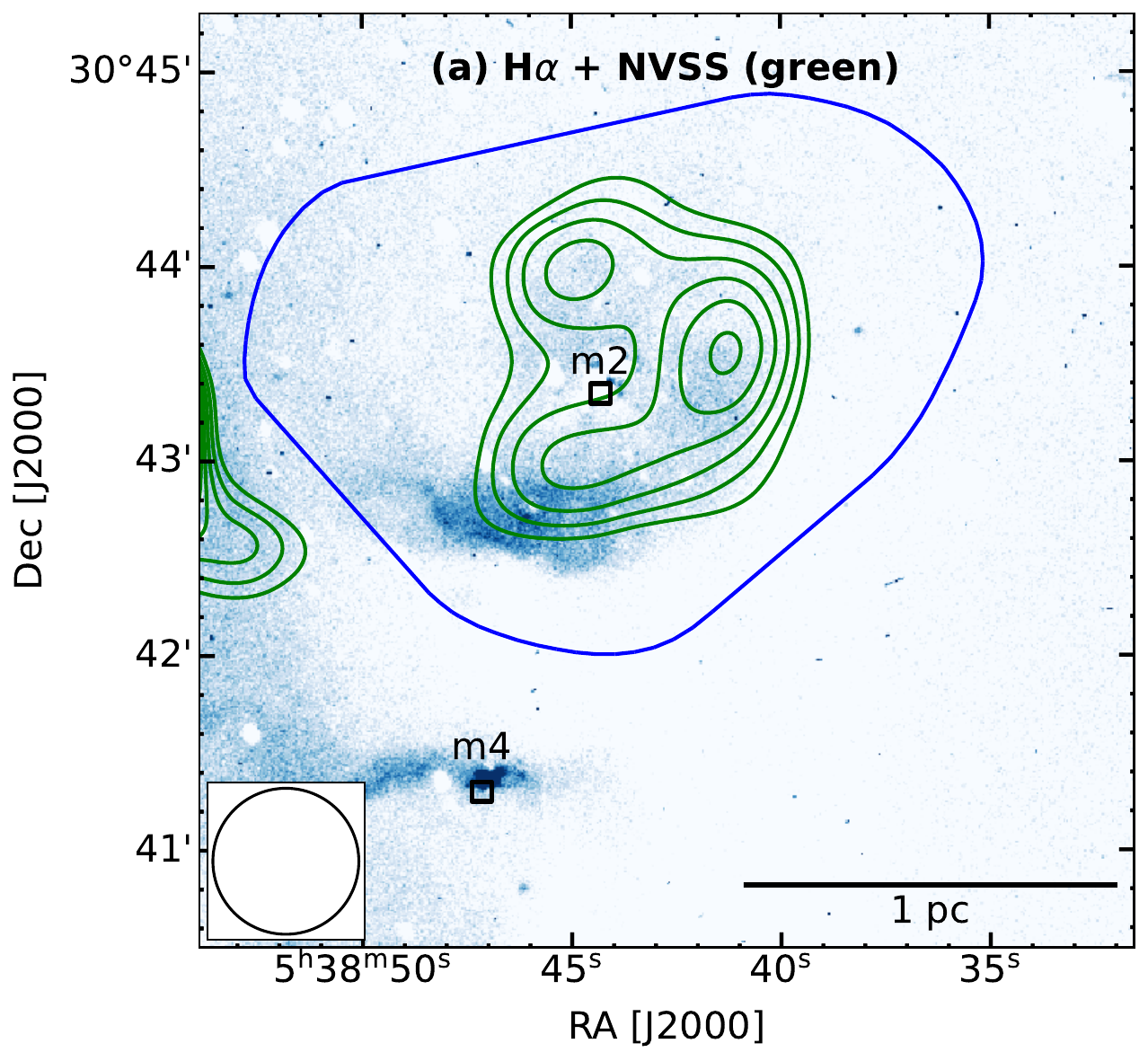}
    \includegraphics[width=0.49\linewidth]{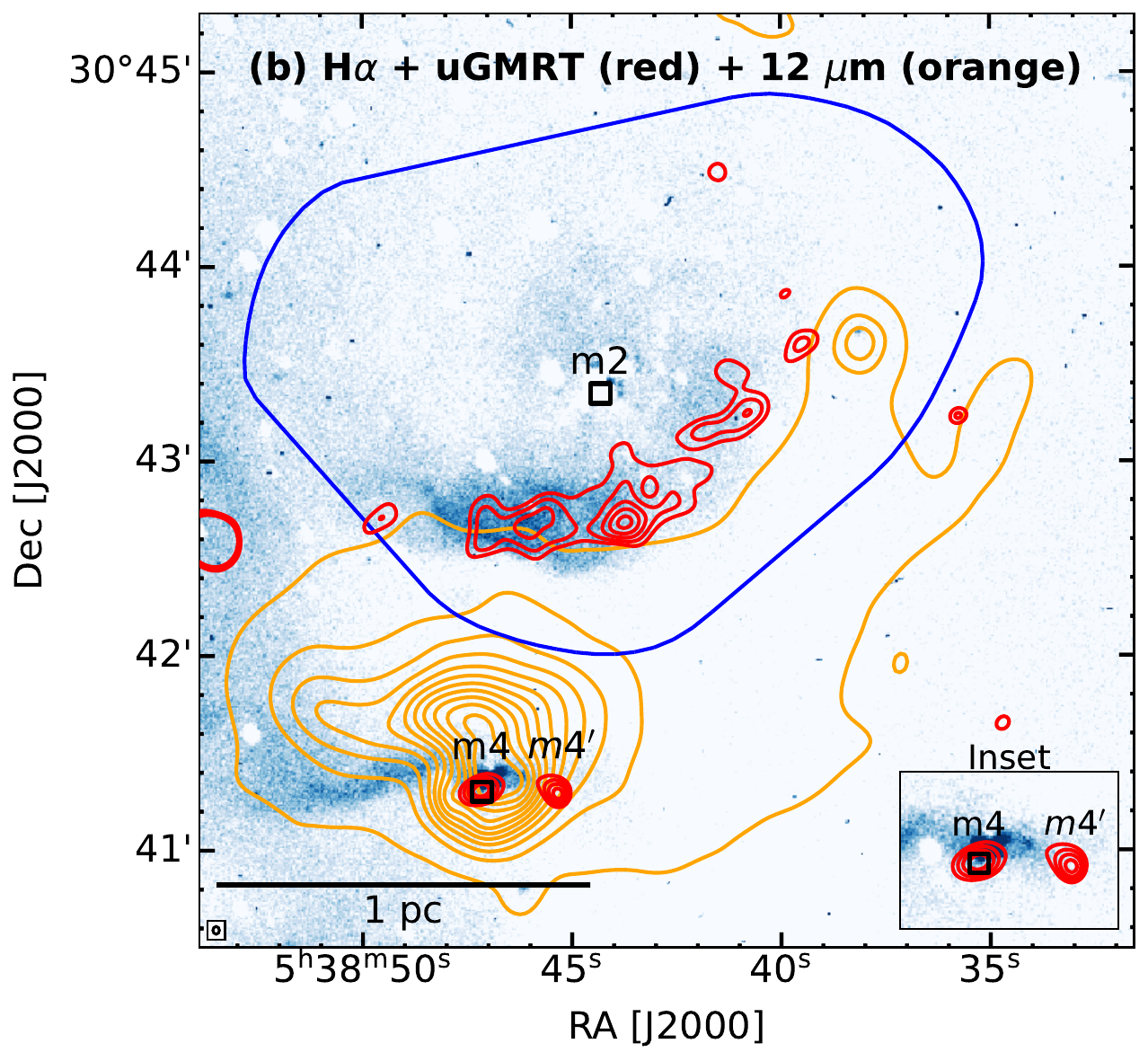}
    \caption{H$\alpha$ emission (using DFOT) to trace the distribution of the ionized gas. It is overlaid with the (a) NVSS 1.4 GHz radio continuum contours (green), the lowest level to generate these contours is 1.50 mJy\,beam$^{-1}$ with a step size of 0.25 mJy\,beam$^{-1}$; and (b) the uGMRT 1260 MHz radio continuum contours (red). The lowest level to generate these contours is 12 $\mu$Jy\,beam$^{-1}$ with a step size of 2.5 $\mu$Jy\,beam$^{-1}$. The WISE 12 $\mu$m intensity contours have been shown with the orange color, with the lowest contour at 1100 counts and the step size of 302 counts. Locations of `m2, m4, and m4$^{\prime}$' are also marked. Both panels are overlaid with a blue polygon representing Cl1 (see Section~\ref{sec:clustering}). The inset in panel (b) represents the zoomed-in view of the radio emission around `m4' and `m4$^{\prime}$' to get a clearer picture. The respective beam shapes of NVSS ($\sim45\arcsec$) and uGMRT 1260 MHz ($\sim2.4\arcsec\times2.0\arcsec$) are shown in the bottom left corners of both the panels.}
    \label{fig:gmrt}
\end{figure*}

Figure~\ref{fig:gmrt} depicts the H$\alpha$ emission at 6563 \r{A}
% in an $18\arcmin.5 \times18\arcmin.5$ region (marked with a green square) 
observed using 1.3-m DFOT (see Section~\ref{sec:halpha}). It is an excellent indicator of star formation activities due to its less susceptibility to extinction or metallicity than the other optical emission lines, such as [O {\sc ii}] \citep{2006ApJ...642..775M}. It traces the distribution of the diffused (warm) ionized gas.
We observe an arc of H$\alpha$ emission near `m2' and `m4'. We have also overlaid this map with the NVSS 1.4 GHz radio continuum emission (Figure~\ref{fig:gmrt}(a), green contours). We see some diffuse radio emission which is almost circularly distributed around `m2', inside the stellar clustering Cl1 (shown with blue polygon). This kind of radio emission is a typical feature of an H\,{\sc ii} region formed due to the strong UV feedback from the massive star(s) \citep{2010A&A...518L.101Z,2014A&A...566A.122S}.

The coarse beam size of NVSS ($\sim45\arcsec\times45\arcsec$; \citealt{1998AJ....115.1693C}) limits the identification of small-scale structures, and the finer distribution of the ionized gas remains unidentified. Thus, to figure that out, we plotted the uGMRT 1260 MHz emission contours (beam size $\sim2.4\arcsec\times2.0\arcsec$) on the H$\alpha$ image (shown in Figure~\ref{fig:gmrt}(b) with red contours). We observe the uGMRT radio emission as an arc around `m2' (completely inside the boundary of Cl1, shown with blue polygon). The ionized gas seems to be density-bounded by the PDRs, as indicated by 12 $\mu$m emission (orange contours in Figure~\ref{fig:gmrt}(b)), in the western direction.  

We  estimated the Lyman continuum flux $N_{UV}$ (photons/s) using the following equation \citep{1976AJ.....81..172M}:

\begin{equation}
    \begin{split}
        N_{UV} \, (s^{-1}) = 7.5 \times 10^{46} \times \left[\frac{S_{\nu}}{Jy} \right] \times \left[\frac{D}{kpc} \right]^2 \times \\
        \left[\frac{T_{e}}{10^4} \right]^{-0.45} \times \left[\frac{\nu}{GHz} \right]^{0.1}
    \end{split}
\end{equation}

Here $S_{\nu}$ denotes the integrated flux in $Jy$, $D$ is the distance in kpc of the target region (1.81 kpc for E71), $T_e$ is the electron temperature, and $\nu$ is the frequency of the band in which the source is observed (in MHz). We considered $T_e = 10000 \, K$, considering that all ionizing flux was produced by a single massive OB star. We determined $S_{\nu}$ as $17\pm0.8$ mJy (around `m2') by integrating the flux down to the lowest contour at 3$\sigma$ level ($\sigma\sim5~\mu$Jy/beam is the rms noise). Using these values, we determined $N_{UV}$ as $4.32 \times 10^{45}$ photons\,s$^{-1}$, which is lower than that reported by \citet{2004A&A...427..839C} for a B1.5-type (for e.g, `m2') massive star ($=1.00 \times 10^{46}$ photons\,s$^{-1}$). We believe that the remaining $N_{UV}$ has been absorbed by dust grains before contributing to ionization \citep{2001AJ....122.1788I,2020ApJ...898..172D}.

Our high-resolution and sensitive uGMRT radio observations reveal two distinct radio continuum peaks: one is coincident with `m4' and another is at a location $\sim$25\arcsec\, west of it (this will be referred to as `m4$^{\prime}$' throughout the paper). The radio emission peaks at `m4' and `m4$^{\prime}$' correspond to sources A and B identified by \citet{2021MNRAS.504..338P}, respectively, who proposed that source A is likely an MYSO, while source B is a probable Herbig–Haro object. They also coincide with peaks of MIR emission.

The MYSO phase begins when the protostellar envelope is sufficiently heated to become detectable in the MIR, and it concludes with the formation of an ultracompact\,H\,{\sc ii} region, which results from the ionization of the surrounding medium by the emerging massive star. Although MYSOs are key to understanding early massive star formation, they are difficult to detect at optical and infrared wavelengths due to high extinction and their lack of strong H\,{\sc ii} region signatures. However, in the last few years, several weak and compact radio sources associated with MYSOs have been detected \citep{1994ApJ...421L..51H,2021MNRAS.504..338P}. 
% In contrast, UC\,H\,{\sc ii} regions are radio-loud, making them more readily identifiable.}

We computed the ionizing photon fluxes for `m4' and `m4$^{\prime}$' as $N_{UV} = 5.1 \times 10^{44}$ and $4.0 \times 10^{44}$ photons\,s$^{-1}$, based on measured flux densities of $S_{\nu} = 2.0\pm0.3$ and $1.6\pm0.3$ mJy, respectively. These ionizing photon rates correspond to B2-type stars (considering ZAMS) powering them, consistent with previous classifications \citep{1973AJ.....78..929P}.

We also estimated the spectral index $\alpha$ for `m4' and `m4$^{\prime}$' using flux measurements from our observed uGMRT Band 5 data (mentioned in the previous paragraph) and those reported by \citet{2021MNRAS.504..338P} for VLA C-band (5.8 GHz) data (0.11 and 0.07 mJy for `m4' and `m4$^{\prime}$', respectively). We assumed a power-law dependence of the form $S_{\nu} \propto \nu^{\alpha}$, where $\nu$ is the frequency. The resulting spectral indices are $\alpha = -1.90 \pm 0.12$ for `m4' and $\alpha = -2.05 \pm 0.19$ for `m4$^{\prime}$', respectively. If $\alpha > 0.1$, the regions are generally attributed to thermal emission \citep{1975A&A....39..217O}, whereas $\alpha < -0.5$ typically indicates that non-thermal processes dominate \citep{1999ApJ...527..154K}. This reveals the existence of non-thermal synchrotron emission for both `m4' and `m4$^{\prime}$' and rules out the UC~H\,{\sc ii} scenario. The non-thermal emission results from jets driven by MYSOs, where particles interact with the magnetic fields of the jet or the surrounding medium. Non-thermal radio emissions from MYSOs have also been reported in previous studies \citep{2019MNRAS.486.3664O}. However, the aforementioned values of $\alpha$ should be considered with caution, as they are approximate, and determining the exact spectral index will require low-frequency data along with high-frequency data that share the same UV coverage and beam size.

%%%%%%%%%%%%%%%%%%%%%%%%%%%%%%%%%%%%%%%%

\subsection{Feedback Pressure employed by the Massive Star `m2'}\label{sec:pressure}

The feedback pressure exerted by the massive star on its surroundings has a pivotal role in the self-regulation of active star formation. This pressure consists of three components (see \citealt{2012ApJ...758L..28B}):

\begin{enumerate}
    \item Pressure exerted by H\,{\sc ii} region:
    \begin{equation}
        P_{HII} = \mu m_{H} c_{s}^2\, \left( \sqrt{\frac{3N_{uv}}{4\pi\,\alpha_{B}\, D_{s}^3}} \right),
    \end{equation}
    \item Radiation Pressure:
    \begin{equation}
        P_{rad} = \frac{L_{bol}}{4\pi c D_{s}^2}, 
    \end{equation}
    \\and\\
    \item Ram pressure employed by the stellar wind:
    \begin{equation}
        P_{wind} = \frac{\dot{M}_{w} V_{w}}{4 \pi D_{s}^2}.
    \end{equation}
\end{enumerate}

Here, $\mu$ is the mean molecular weight of the ionized gas ($=$ 0.678; \citealt{2009A&A...497..649B}), $m_H$ is the atomic mass of hydrogen, $c_s$ is the sound speed in the photoionized region ($=11$ \kms; \citealt{2004fost.book.....S}), $N_{uv}$ is the Lyman continuum photons, $\alpha_{B}$ is radiative recombination coefficient ($= 2.6 \times 10^{-13} \times (10^4 K / T_e)^{0.7}$ cm$^3$\,s$^{-1}$, \citealt{1997ApJ...489..284K}), $D_s$ is the projected distance between the massive star and the location at which the pressure is to be calculated, $L_{bol}$ is the bolometric luminosity, $\dot{M}_{w}$ denotes mass-loss rate and $V_{w}$ is wind velocity of the ionizing source.

For B1.5V type star,  the values for $N_{uv} = 1.00 \times 10^{46}$ photons\,s$^{-1}$ \citep{2004A&A...427..839C}, $\dot{M}_{w} = 1.02 \times 10^{-8}$ M$_{\odot}$\,yr$^{-1}$ \citep{2023A&A...678A.172P}, $V_w =$ 1200 \kms \citep{2023A&A...678A.172P} and
$L_{bol}$ = 9332.54 L$_{\odot}$ \citep{2023A&A...678A.172P}. 

We took $D_s =$ 1.1 pc as the projected distance between `m2' and MYSO `m4'. These values yield $P_{HII} = 2.1 \times 10^{-11}$, $P_{rad} = 8.1 \times 10^{-12}$, $P_{wind} = 5.2 \times 10^{-13}$ dynes\,cm$^{-2}$ and thus total pressure ($=P_{HII} + P_{rad} + P_{wind}$) exerted by `m2' on `m4' is $2.9 \times 10^{-11}$ dynes\,cm$^{-2}$. Since `m2' is located almost at the center of the bubble, the pressure exerted by `m2' on the periphery of the E71 bubble will approximately be equal to the pressure exerted by `m2' on `m4'.

%%%%%%%%%%%%%%%%%%%%%%%%%%%%%%%%%%%%%%%%%

\subsection{Molecular (CO) Morphology}
\label{sec:MolecularMorphology}

\subsubsection{Dynamics and Morphology of the Molecular Gas}
\label{sec:ChannelMap}

\begin{figure*}[!ht]
    \centering
    \includegraphics[width=\textwidth]{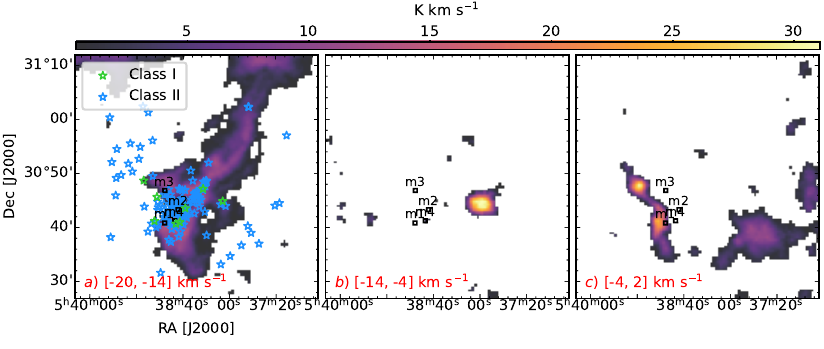}
    \caption{Integrated-intensity map for the \co\, emission over the velocity intervals (in \kms) mentioned in each panel. The emission is shown above 5$\sigma$ value ($\sigma$ being the rms noise). All the panels are overlaid with the locations of probable massive stars (black squares), and panel (a) is marked with the locations of classified YSOs (Class $\textsc{i}$ with green asterisks, and Class $\textsc{ii}$ with blue asterisks).}
    \label{fig:12co_channel_maps1}
\end{figure*}

This section examines the molecular gas dynamics and morphology around the E71 bubble. We selected a large-scale view around the bubble to know the morphology of the local molecular cloud. Using the PMO CO data, we found the existence of two molecular clouds in the velocity range [$-20$, $-14$] \kms\, and [$-4$, $2$] \kms\, in this region (see Figure~\ref{fig:12co_channel_maps1}). We have also found a molecular clump in the intermediate velocity range [$-12$, $-8$] \kms.

On inspecting the integrated-intensity maps, we noticed that the molecular cloud in the velocity range [$-20$, $-14$] \kms\, has a bubble morphology and seems to be associated with the E71 bubble. This morphology is also associated with a filamentary structure extending towards the northwest direction. 
We further analyze the dynamics and morphology of the molecular cloud, particularly focusing on [$-20$, $-14$] \kms\, velocity range, in the subsequent sections.

\subsubsection{Moment Maps}
\label{sec:MomentMaps}

\begin{figure*}[!ht]
    \centering
    \includegraphics[width=\textwidth]{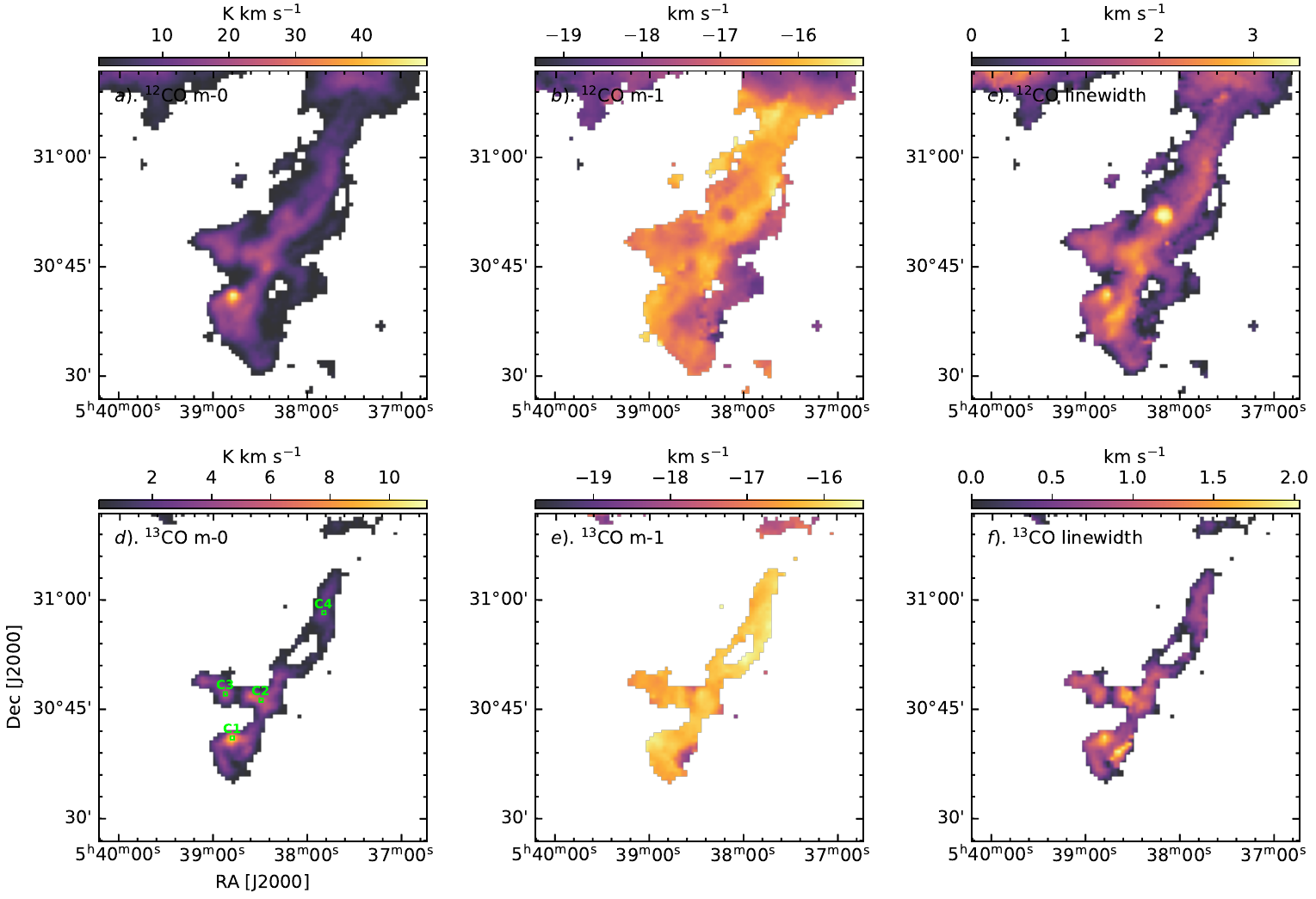}
    \caption{m-0, m-1 and linewidth maps (column-wise) for \co\, and \tco\, emission, respectively in the velocity range [$-20$, $-14$] \kms . The emission is depicted above 5$\sigma$ value ($\sigma$ being the rms noise for the respective spectral cubes). The green squares in panel (d) mark the location of the local peak emission.}
    \label{fig:moment_maps}
\end{figure*}

We have shown the \co\, and \tco\, Integrated Intensity (moment-0 or m-0), Intensity-weighted velocity (moment-1 or m-1), and Intensity-weighted dispersion (linewidth) maps for the E71 bubble in Figure~\ref{fig:moment_maps}. The \co\, and \tco\, m-0 maps (panels (a) and (d) of Figure~\ref{fig:moment_maps}, respectively) 
% in the velocity range [$-20$, $-14$] \kms, 
confirms the bubble-like structure at E71, which is associated with a larger filamentary structure. The location of `m4' has the maximum emission. The m-1 maps (panels (b) and (e) of Figure~\ref{fig:moment_maps}) reveal that the entire bubble is almost at uniform velocity ($-16$ \kms). The
linewidth maps (panels (c) and (f) of Figure~\ref{fig:moment_maps}) for both transitions show maxima towards the arc of the bubble, suggesting gas flows and/or outflow activities in it.
%towards the MYSO `m4'. 

\subsubsection{Position-Velocity (PV) Diagrams}
\label{sec:pvdiagram}
\begin{figure*}[!ht]
    \centering
    \includegraphics[width=0.35\textwidth]{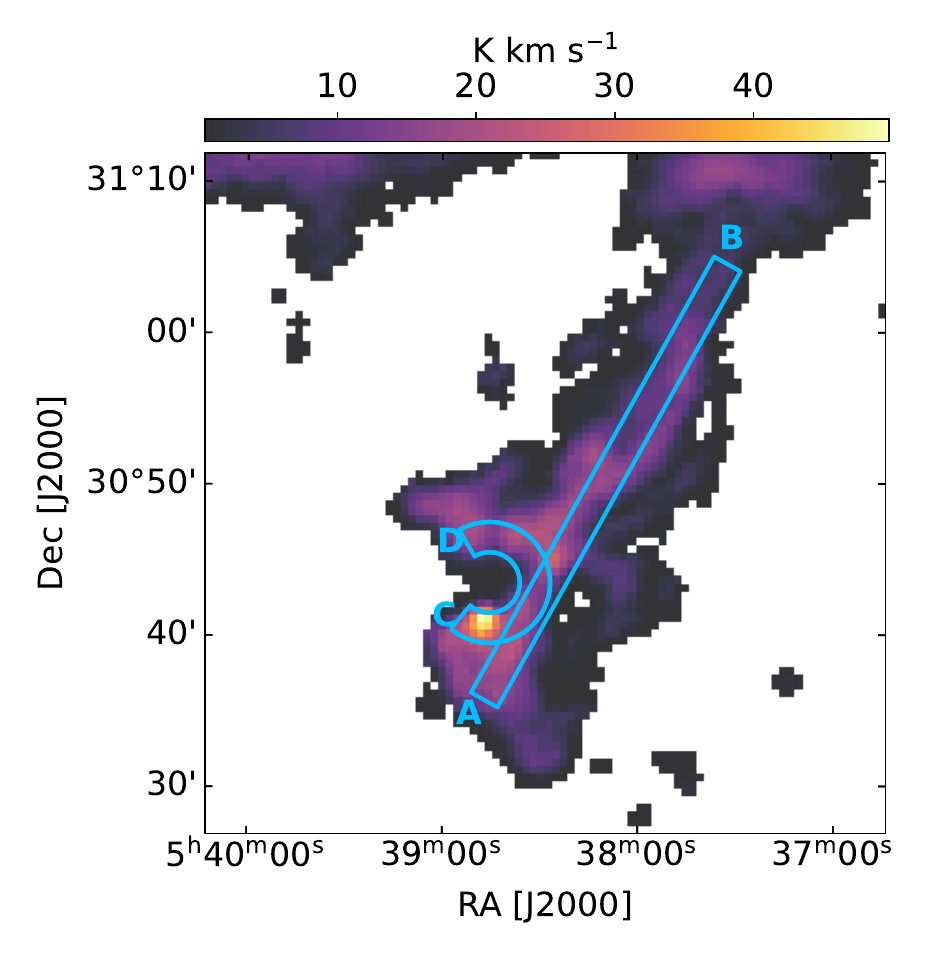}
    \includegraphics[width=0.37\textwidth]{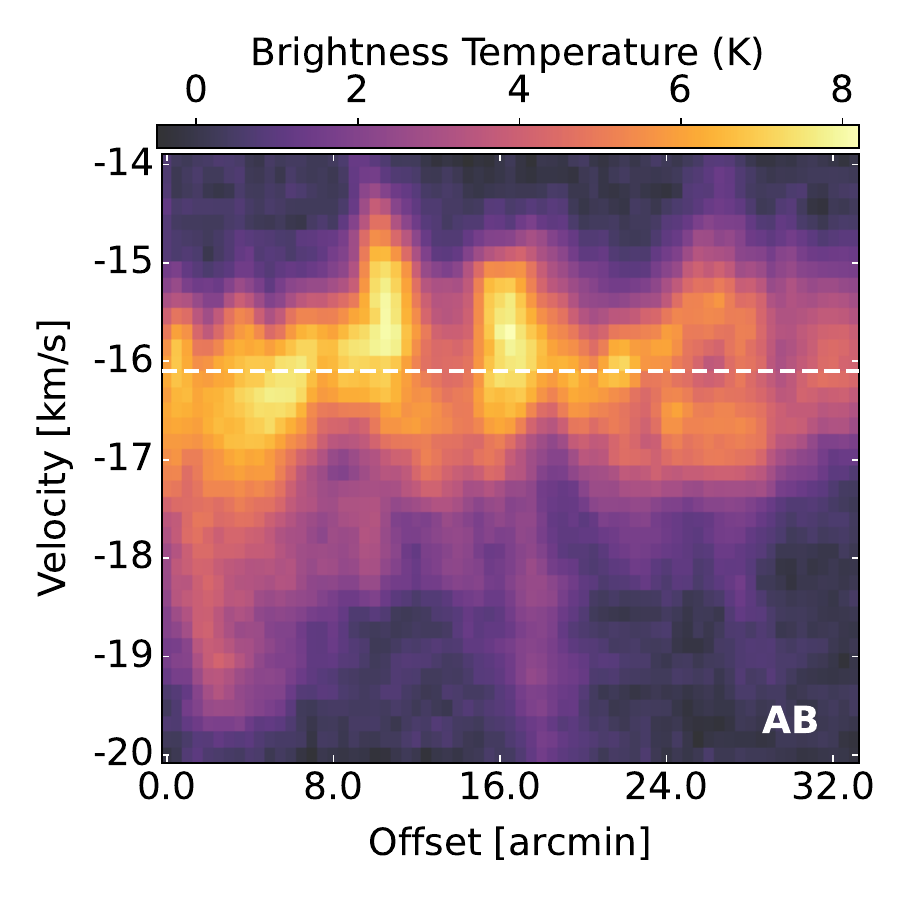}
    \includegraphics[width=0.25\textwidth]{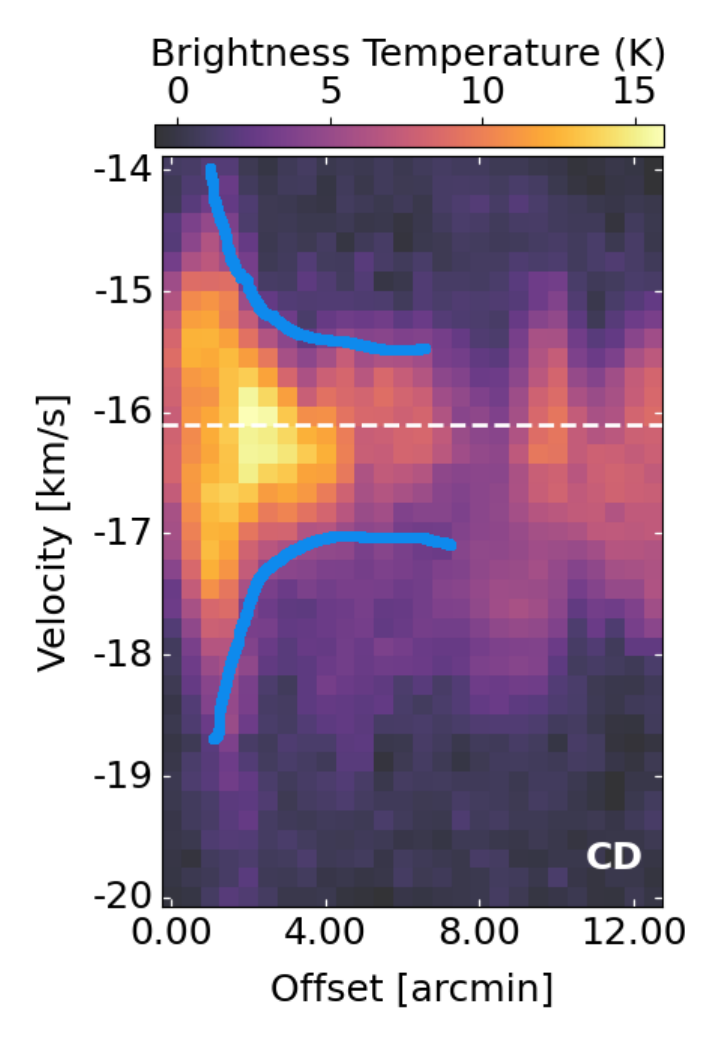}
    \caption{Left Panel: \co\, m-0 map overlaid with elected paths AB and CD to extract the PV maps. Middle and right panels: PV Diagrams extracted using \co\, along AB and CD, respectively. The blue curves in the right panel represent the expansion of the E71 bubble.}
    \label{fig:pv_diagrams}
\end{figure*}

We observed in the moment maps (Figure~\ref{fig:moment_maps}) that the E71 bubble is adjacent to a filamentary structure, though there is a significant fluctuation in velocity. To gain a better understanding of these fluctuations, we generated position-velocity (PV) maps along the paths AB and CD (shown in the left panel of Figure~\ref{fig:pv_diagrams}) using the \co\, spectral data cube. The thought behind electing these paths is that we want to see the velocity variation along the (filament $+$ bubble) and the bubble. While inspecting the PV map along the path AB (middle panel of Figure~\ref{fig:pv_diagrams}), we observed a velocity spread along both sides of $\sim-16$ \kms, this spread is more ($\sim$3 \kms) towards A, i.e., towards the bubble. The PV maps along the arc CD (right panel of Figure~\ref{fig:pv_diagrams}) reveal that the bubble structure has a velocity spread along both sides of $\sim-16$ \kms, suggesting its expanding nature. The mean velocity along arc CD is approximately $-16$ \kms, with a spread over [$-18$, $-14$] \kms\, (see right panel of Figure~\ref{fig:pv_diagrams}), yielding an inferred expansion velocity of $\sim$2 \kms.

Along the arc CD, a molecular clump is detected at $\sim2\arcmin$, coinciding with the location of the MYSO `m4'. This clump exhibits a brightness temperature of approximately 15.5\,K. Notably, this location also hosts two probable UC H\,{\sc ii} regions. In the linewidth map, this region displays a higher velocity dispersion compared to the rest of the bubble’s periphery (panels (c) and (f) of Figure~\ref{fig:moment_maps}). This enhanced velocity dispersion may be attributed to strong outflow activities associated with the embedded MYSO and the adjacent UC H\,{\sc ii} regions.

%%%%%%%%%%%%%%%%%%%%%%%%%%%%%%%%%%%%%%%%

\subsubsection{Physical Parameters}

% \begin{figure*}[!ht]
%     \centering
%     \includegraphics[width=\textwidth]{fig/13co_spectra_all.pdf}
%     \caption{\tco\, (blue) spectra at local peaks of m-0 emission (C1, C2, C3, C4, and C5; marked in panel (d) of Figure~\ref{fig:moment_maps} with green squares), IRAS sources (I1, I2, I3, and I4; marked with red diamonds in the upper left panel of Figure~\ref{fig:intro}), and MSX sources (S1, S2, S3, and S4; marked with red triangles in the lower right panel of Figure~\ref{fig:intro}). The green curves represent the Gaussian fit/s to these spectra. The vertical dashed (blue) lines represent the $^{12}$CO Gaussian fit/s peak, which are also mentioned in the respective sub-panels.}
%     \label{fig:co_spectra}
% \end{figure*}

We derived various parameters using the Gaussian model fitting on the non-averaged \tco\, spectral cube in the velocity range [$-20$, $-14$] \kms\, 
% (see Figure~\ref{fig:co_spectra}) 
using the following set of equations \citep{1992ApJ...384..523F,2000MNRAS.311...85F}:

\begin{equation}
    \begin{split}
        \Delta V^2_{tot} = \Delta V^2_{obs} + 8 \ln 2 \, kT \left( \frac{1}{\bar{m}} - \frac{1}{m_{obs}} \right) \\
        \Rightarrow \frac{\Delta V^2_{tot}}{8 \ln 2} = \frac{kT}{\bar{m}} + \left( \frac{\Delta V^2_{obs}}{8 \ln 2} - \frac{kT}{m_{obs}} \right)\\
        \Rightarrow \sigma^2_{tot} =  c^2_s + (\sigma^2_{obs}-\sigma^2_{T})\\
        = c^2_s + \sigma^2_{NT}
    \end{split}
\end{equation}

and

\begin{equation}
    M_{line,virial} = \left[1 + \left( \frac{ \sigma_{NT}}{c_s} \right)^2 \right] \times \left[16 M_\odot\,pc^{-1} \times \left(\frac{T}{10 K} \right) \right]
\end{equation}

Here, $V_{obs}$ and $\sigma_{obs}$ ($=V_{obs}/\sqrt{8\,ln2}$) are the full width at half maximum (FWHM) and velocity dispersion (standard deviation), respectively, of the Gaussian fit of the obtained spectrum; $\sigma_{T}$ ($=\sqrt{kT/m_{obs}}$) and $\sigma_{NT}$ are the thermal and non-thermal velocity dispersion, respectively; $T$ is the excitation/kinetic temperature; $c_s$ ($=\sqrt{kT/\bar{m}}$) is the speed of sound; $m_{obs}$ is the mass of the \tco\, molecule (29 amu); and $\bar{m}$ is the mean molecular weight of the medium (2.37 amu); and $M_{line,virial}$ is the virial line mass. Assuming that the dust and gas temperatures are coupled via collisions \citep{2001ApJ...557..736G,2023ApJ...944..228M}, we used $T_d$ from the \textit{Herschel} temperature map as the gas kinetic temperature $T$. However, a substantial difference between $T_d$ and $T$ may still exist under certain physical conditions. We have mentioned the derived parameters at local peaks of \tco\, m-0 emission (C1, C2, C3, and C4; marked in panel (d) of Figure~\ref{fig:moment_maps} with green squares), IRAS sources (I1, I2, and I3; marked with red diamonds in the upper left panel of Figure~\ref{fig:intro}), and MSX sources (S1, S2, S3, and S4; marked with red triangles in the lower right panel of Figure~\ref{fig:intro}) in the Table~\ref{tab:co_parameters}. The spectra have been extracted at these locations for square-shaped regions of size $30\arcsec \times 30\arcsec$. Though there are more IRAS and MSX sources, we selected only those that do not overlap with each other and are related to the bubble and the filamentary structure. We used the above-estimated values also to calculate the ratio of thermal-to-non-thermal pressure ($P_{TNT} = c_s^2/\sigma_{NT}^2$) and the Mach number ($=\sigma_{NT}/c_s$) \citep{2003ApJ...586..286L}. $P_{TNT}$ reveals the dominance of the non-thermal pressure component over the thermal pressure component in the molecular cloud, whereas the Mach number indicates the presence of supersonic motion. Thus, these values indicate supersonic non-thermal emission phenomena.

\begin{table*}[!ht]
    \footnotesize
   \centering
   \caption{Physical parameters derived using the \tco\, molecular line data.}
   \begin{tabular}{c c c c c c c c c c c}
   \hline
    Regions & $\alpha_{J2000}$ & $\delta_{J2000}$ & FWHM & T & C$_s$ & $\sigma_{T}$ & $\sigma_{NT}$ & P$_{TNT}$ & Mach & M$_{line,vir}$\\
     & (hh:mm:ss) & (dd:mm:ss) & (\kms) & (K) & (\kms) & (\kms) & (\kms) & & Number &  (M$_\odot$/pc)\\
    \hline
    Local Peaks of m-0 Emission:\\
    C1 & 05:38:48 & +30:41:11 & 1.99 & 19.96 & 0.26 & 0.08 & 0.85 & 0.10 & 3.18 & 356\\
    C2 & 05:38:27 & +30:44:41 & 1.51 & 14.75 & 0.26 & 0.08 & 0.64 & 0.17 & 2.42 & 217\\
    C3 & 05:38:50 & +30:47:10 & 1.52 & 14.68 & 0.23 & 0.06 & 0.65 & 0.12 & 2.84 & 212\\
    C4 & 05:38:13 & +30:51:40 & 1.08 & 14.00 & 0.22 & 0.06 & 0.46 & 0.24 & 2.05 & 117\\
    \hline
    IRAS Sources:\\
    I1 & 05:38:38 & +30:42:51 & 1.69 & 18.43 & 0.25 & 0.07 & 0.72 & 0.13 & 2.81 & 262\\
    I2 & 05:38:44 & +30:45:36 & 0.97 & 16.85 & 0.24 & 0.07 & 0.41 & 0.36 & 1.66 & 101\\
    I3 & 05:38:02 & +30:53:20 & 0.62 & 14.04 & 0.22 & 0.06 & 0.26 & 0.76 & 1.15 & 52\\
    \hline
    MSX Sources:\\
    S1 & 05:38:40 & +30:41:29 & 0.90 & 17.67 & 0.25 & 0.07 & 0.38 & 0.44 & 1.51 & 92\\
    S2 & 05:38:40 & +30:42:23 & 1.11 & 18.43 & 0.25 & 0.07 & 0.47 & 0.30 & 1.84 & 129\\
    S3 & 05:38:38 & +30:44:00 & 0.96 & 17.85 & 0.25 & 0.07 & 0.41 & 0.39 & 1.61 & 102\\
    S4 & 05:38:41 & +30:46:00 & 1.10 & 15.52 & 0.23 & 0.07 & 0.47 & 0.25 & 1.98 & 122\\
    \hline\\
   \end{tabular}
    
   \label{tab:co_parameters}
\end{table*}

We also determined the mass of the filamentary structure adjacent to the E71 bubble (traced by path AB in Figure~\ref{fig:pv_diagrams}). 
% To estimate its mass, we converted the \co\ integrated intensity ($I_{^{12}CO}$) in the [$-20$,$-14$] \kms\, range into H$_2$ column density using the relation $N{\rm H_2} = X_{\rm ^{12}CO} \cdot I_{^{12}CO}$ \citep{2020MNRAS.497.1851S}, adopting the standard conversion factor $X_{\rm ^{12}CO} = 2.0 \times 10^{20}$ cm$^{-2}$\,(K\,km\,s$^{-1}$)$^{-1}$. The total mass of the filament is estimated to be $\sim$973 M$_\odot$.
We converted the \tco\ integrated intensity ($I_{^{13}CO}$) in the [$-20$,$-14$] \kms\, range into H$_2$ column density by considering the local thermal equilibrium (LTE) assumption using the relation \citep{2020MNRAS.497.1851S}, 

\textbf{\begin{equation}
    N_{H_2}(^{13}CO) = X_{^{13}CO,Q}\,I_{^{13}CO}
\end{equation}
}
where

\textbf{\begin{equation}
    X_{^{13}CO,Q} = 1.5 \times 10^{20}\,Q\,[H_2\,cm^{-2}(K\,km\,s^{-1})^{-1}]
\end{equation}}
is the conversion factor for \tco\, transition. The factor $Q$ is given as

\textbf{\begin{equation}
    Q = Q(T_{ex}, T_{B}(^{13}CO))=\frac{\tau}{1-e^{-\tau}}\,\frac{1}{1-e^{-T_0^{110}/T_{ex}}}.
\end{equation}}

Here $T_{ex}$ is the excitation temperature, be estimated as,
\textbf{\begin{equation}
    T_{ex} = T_0^{115}\times ln \left( 1 + \frac{T_0^{115}}{T_B(^{12}CO)_{max}+0.83632} \right)^{-1}\, K
\end{equation}}

and $\tau$ is the optical depth for \tco\, transition, estimated as,

\textbf{\begin{equation}
    \tau = - ln \left( 1 - \frac{T_B(^{13}CO)_{max}/T_0^{110}}{(e^{T_0^{110}/T_{ex}}-1)^{-1}-0.167667} \right)^{-1}\, K.
\end{equation}}

Here, $T_0^{115}$ and $T_0^{110}$ are the Planck temperatures, $T_0=h\nu/k$, at corresponding frequencies, and $T_B$ is the main beam temperature.

The total mass of the filament is estimated to be $\sim$116~M$_\odot$.

We also determined the masses of the molecular condensations in the red-shifted ([$-4$, $2$] \kms) and the blue-shifted ([$-20$, $-14$] \kms) molecular clouds (marked with yellow color in the right panel of Figure~\ref{fig:intro}), considering their morphology as circular/elliptical (sizes are mentioned in Table~\ref{tab:molecular_condensations}), using the aforementioned method, and the obtained the values of the total mass of the core, as shown in Table~\ref{tab:molecular_condensations}.

 We evaluated the number density of the cores ($n(H_2)$), which is defined as the ratio of their total mass to their volume \citep{2005fost.book.....S}, i.e., 

\textbf{\begin{equation}
    n(H_2) = \frac{M_{core}}{V_{core}\,\mu \, m_H} = \frac{M_{core}}{(\frac{4}{3}\, \pi \, a\,b\,c)\,\mu \, m_H}.
\end{equation}}

Here, $M_{core}$, $R_{core}$, and $V_{core}$ represent the mass, radius, and volume of the core, respectively. $\mu=2.8$ is the mean molecular weight \citep{2008A&A...487..993K}, and $m_H$ is the atomic mass of the Hydrogen atom. 
% We considered $R_{core}$ as the geometric radius of the core given as $R_c=\sqrt{ab}$, where 
The variables $a$, $b$, and $c$ represent the lengths of the semi-axes corresponding to the three axes of the cores. As we did not have specific information regarding the length of the third axis, we typically considered $c$ to be the average of the semi-major and semi-minor axes. The calculated number densities are listed in Table~\ref{tab:molecular_condensations}. These values align with the typical number densities found in dense cores \citep{2005fds..book.....S}.

\begin{table*}[!ht]
    \centering
    \caption{Various parameters for the molecular condensations marked in the right panel of Figure~\ref{fig:intro}}
    \begin{tabular}{c c c c c c c}
    \hline
    Name & $\alpha_{J2000}$ & $\delta_{J2000}$ & Size (Major $\times$ Minor Axes) & Total Mass & Number Density\\
     & (hh:mm:ss) & (dd:mm:ss) & (pc $\times$ pc) & ($M_\odot$) & ($\times 10^3\,cm^{-3}$)\\
    \hline
    B1 & 05:38:48 & +30:40:37 & 1.02$ \times $1.02 & 76.8 & 2.0\\
    B2 & 05:38:27 & +30:45:49 & 0.57$ \times $0.44 & 12.5 & 2.7\\
    B3 & 05:38:52 & +30:47:33 & 1.01$ \times $0.65 & 22.5 & 1.1\\
    R1 & 05:39:19 & +30:47:30 & 0.76$ \times $0.76 & 20.54 & 1.2\\
    R2 & 05:39:02 & +30:40:53 & 1.14$ \times $0.90 & 42.2 & 1.1\\
    \hline\\
    \end{tabular}
   
    \label{tab:molecular_condensations}
\end{table*}

%%%%%%%%%%%%%%%%%%%%%%%%%%%%%%%%%%%%%%%%

\section{Discussion}\label{sec:discussion}
The E71 bubble and its surrounding region exhibit multiple signatures of recent and ongoing star formation, including the presence of Class\, {\sc i} YSOs, young and massive stars, PDRs, warm and cold gas and dust, ionized gas, and molecular clouds. These characteristics collectively highlight E71 as a compelling and active site of star formation. To investigate the ongoing star formation activity in this region, we employed a comprehensive multi-wavelength approach.

This panel also includes an overlay of Class\,\textsc{i} YSOs to examine their spatial distribution relative to the molecular gas. A clear association is observed, suggesting that the target region is actively forming stars.

\subsection{The stellar cluster and E71 bubble}\label{sec:sf_scenario}

We first examined the stellar density distribution using deep NIR data and identified a stellar cluster, Cl1, situated within the E71 bubble. This cluster has an estimated radius of $\sim$1.26 pc and hosts a massive star of spectral type B1.5. Based on this classification, we estimate an upper age limit of approximately 12 Myr for Cl1. Using a probabilistic approach with \textit{Gaia} DR3 parallax data, we determine the cluster's distance to be $\sim$$1.81 \pm 0.15$ kpc. The massive star `m2' is confirmed as a member of Cl1.

We also derived the $\Gamma$ through least-square fitting for Cl1 and found that it has a break in the $\Gamma$ at $\sim$0.7 M$\odot$. We calculated that $\Gamma=-1.98 \pm 0.33$ and $+1.58 \pm 0.47$ within the mass ranges $\rm \sim0.7 < M/M_\odot < 5.2$ and $\rm \sim0.2 < M/M_\odot < 0.7$, respectively, a trend reported in several previous studies of young clusters.

MIR/FIR observations reveal an arc or ring of gas and dust encircling Cl1 and regularly spaced molecular and dust condensations. The arc/ring-like structure is also evident in our extinction and stellar density maps, which indicate that Cl1 is surrounded by a high-extinction environment. Moreover, the spatial overlap between the \textit{convex hull} of the YSO core and regions of enhanced column density suggests that new star formation is occurring within these dense regions.

Analyses of WISE 12 $\mu$m emission, \textit{Spitzer} IRAC band ratio maps (4.5 $\mu$m/3.6 $\mu$m), NVSS 1.4 GHz radio continuum data, and \textit{Herschel} intensity and column density maps reveal a lotus-shaped PDR centered around Cl1 and `m2'. The ionized gas within this cavity appears to be density-bounded. Elevated dust temperatures along the arc further support the influence of stellar feedback in shaping the environment. 

Interestingly, this high-extinction arc also hosts a deeply embedded massive MYSO, designated as `m4'
% , along with a probable UC H\,{\sc ii} region `m4$\prime$.
These features point to active high-mass star formation, likely influenced by the feedback from Cl1. High-resolution uGMRT radio observations reveal two distinct radio continuum peaks: `m4 and m4$^{\prime}$' \citep[see also][]{2021MNRAS.504..338P}, supporting their identification as candidate UC H\,{\sc ii} regions powered by B3-type stars.

Given that Cl1 lies near the center of the E71 bubble and that `m2' is its only massive member, we propose that `m2' is the primary ionizing source driving the expansion of the bubble. To quantify its influence, we calculated the pressure exerted by `m2' on the surrounding medium. At the location of MYSO `m4', this pressure is estimated to be $2.9 \times 10^{-11}$ dynes\,cm$^{-2}$. This exceeds the internal pressure of a typical molecular cloud, which is $\sim 10^{-11}$–$10^{-12}$ dynes\,cm$^{-2}$, assuming a particle density of $\sim 10^3$–$10^4$ cm$^{-3}$ and a temperature of 20 K \citep[see Table 2.3 of][]{1980pim..book.....D}. Thus, the feedback from `m2' is likely sufficient to compress the surrounding gas and trigger star formation.

We therefore infer that the formation of MYSO `m4' may have been triggered by the radiative and mechanical feedback from `m2'. This supports a scenario in which Cl1 formed first, and subsequent feedback from its massive member(s)—particularly `m2'—sculpted the E71 bubble and initiated a new generation of star formation at its periphery. According to \citet{2005A&A...433..565D}, the ``collect and collapse'' process typically produces a compressed shell that can be observed at MIR, submillimeter, and millimeter wavelengths, with massive clumps or cores distributed at regular intervals along the shell. Our findings align well with their results. 
Similar feedback-driven processes have been reported in other Galactic bubbles \citep[e.g.,][]{2017MNRAS.467.2943S, 2023ApJ...953..145V}, consistent with the ``collect and collapse'' mechanism \citep{1977ApJ...214..725E, 1998ASPC..148..150E}.

\subsection{YSO distribution in the region}

Using the deep NIR photometric catalog, we have identified a total of 8 Class\,{\sc i} and 139 Class\,{\sc ii} YSOs in the E71 region. These YSOs are spatially distributed across the entire E71 bubble and its surrounding area, correlating well with the observed MIR emission and regions of molecular condensation. Notably, the majority of Class\,{\sc i} YSOs are concentrated along the arc of gas and dust that delineates the E71 bubble, suggesting that this structure is currently an active site of star formation. On the other hand, YSOs do not exist along the filamentary cloud associated with the E71 bubble. We looked for the multitemperature maps of the differential column density that help to interpret more complex systems by distinguishing various physical components along the line of sight \citep{2015MNRAS.454.4282M}. We find that the filamentary structure associated with the E71 bubble appears for temperatures below $\rm \sim 16 \,K$ only, and the non-existence of the YSOs indicates that no star formation has taken place yet, and it is at an early stage of fragmentation. We infer that the formation of filaments occurs before star formation in the cold ISM and is related to processes occurring within the clouds themselves.

Using MST analysis, we were able to isolate two prominent groupings from the extended YSO distribution: a dense YSO core and a more spatially extended aggregate region (AR). The core appears to be located in the inner region of the E71 bubble, encompassing both the Cl1 cluster and the surrounding gas and dust arc, highlighting a localized region of heightened star formation activity. In contrast, the AR spans a broader area beyond the bubble.

We estimated various structural parameters and star formation indicators for these regions, summarized in Table \ref{tab:mst}. Our analysis shows that both the AR and core are larger and more elongated than Cl1, with the AR exhibiting a slightly higher fraction of Class\,{\sc i} YSOs. Cl1, on the other hand, contains the lowest fraction of Class\,{\sc i} YSOs, indicative of its relatively older age compared to the other two regions. Furthermore, the YSOs in Cl1 are more tightly grouped, pointing to a more gravitationally bound system having the same star formation history.

In both the AR and the core, the mean separation between YSOs ($\sim$0.33 and 0.26 pc, respectively) is less than the local $\lambda_J$ ($\sim$1.61 and 1.35 pc, respectively), implying the possibility of non-thermally driven fragmentation \citep{2014MNRAS.439.3719C, 2024AJ....168...98V}. Non-thermal fragmentation describes the process by which molecular clouds break into smaller pieces due to influences beyond just thermal pressure, such as turbulence or magnetic fields. These non-thermal forces play a crucial role in controlling how the cloud fragments and eventually lead to star formation activities.

%To assess the spatial distribution of YSOs, we computed the $Q$ parameter. For the AR, %we obtained $Q\sim0.8$, suggesting a nearly homogeneous distribution, consistent with a %region transitioning from a fractal (substructured) to a radial (centrally concentrated) %configuration \citep{2017MNRAS.467..512S}. In contrast, the core yielded $Q = 0.68$, indicating a fractal or substructured morphology. According to %\citet{2004MNRAS.348..589C}, values of $Q > 0.8$ denote radial distributions, while $Q < %0.8$ implies fractal distributions. The evolution of $Q$ as reported by %\citet{2017MNRAS.467..512S} suggests that star formation typically begins in overdense, %fractal gas distributions, which are gradually shaped into more uniform configurations %through stellar feedback processes.
%Hence, the AR appears to be undergoing this transition, likely influenced by the %feedback from massive stars \citep{2010LNEA....4....1B, 2017MNRAS.467..512S}.

We also evaluated the SFE for the three regions. The core exhibits higher SFE than the AR, while Cl1 shows the highest SFE among them. This is possibly due to Cl1's more evolved nature and the effects of stellar feedback, such as winds and radiation, that may have cleared away gas. Interestingly, we also find no clear correlation between SFE and the fraction of Class\,{\sc i} YSOs. This lack of correlation is consistent with earlier studies \citep{2014MNRAS.439.3719C, 2020ApJ...891...81P}, suggesting that while SFE can indicate how efficiently gas is converted into stars, it may not directly reflect the current proportion of the youngest stellar population.

\subsection{Molecular gas distribution and kinematics in the region}\label{sec:mol_gas_distri}

We investigated the molecular gas morphology surrounding the E71 bubble using high-resolution and sensitive CO observations from PMO. Two distinct molecular cloud components were identified in the velocity intervals [$-20$, $-14$] \kms\, and [$-4$, $2$] \kms\, within the observed field. The cloud in the [$-20$, $-14$] \kms\, range appears to host an arc-like structure (denoted as arc CD in the left panel of Figure~\ref{fig:pv_diagrams}), corresponding to the E71 bubble, with peak \co\ and \tco\ emission coinciding with the location of MYSO `m4' and probable UC H\, {\sc ii} regions. This association is further supported by the m-0, m-1, and m-2 moment maps of both \co\ and \tco.

Additionally, we identified a filamentary structure
adjacent to the E71 bubble (traced by path AB in
Figure~\ref{fig:pv_diagrams}). This filament is characterized by a low dust temperature ($T_d \lesssim 16$ K; see Figure~\ref{fig:diffcdens}) and a lack of associated YSOs. This suggests that it is a cold, quiescent structure, likely below the critical density threshold required for gravitational collapse and subsequent star formation. We therefore classify it as a non-star-forming filament.

We also examined the kinematics along the E71 bubble and the adjacent filament. A higher velocity dispersion is observed along the bubble, indicative of molecular gas expansion (with expansion velocity  $\sim$2 \kms) likely driven by the embedded H\,{\sc ii} region. However, there is an important point to be considered. The right panel of Figure~\ref{fig:intro} clearly shows that the red-shifted molecular cloud (ranging from [$-4$, $2$] \kms) and blue-shifted molecular cloud (ranging from [$-20$, $-14$] \kms) appear to form a disrupted molecular shell. In this scenario, the blue and red molecular clouds correspond to components that are approaching/receding. This suggests an expansion velocity of approximately 11 \kms. Nevertheless, since the blue-shifted cloud aligns more closely in morphology with the E71 bubble, we have chosen to focus our analysis on that cloud alone.

To further probe the gas dynamics, we computed the ratio of thermal-to-nonthermal pressure ($\rm P_{TNT}$) and Mach numbers at several locations within the E71 bubble. These calculations confirm the presence of supersonic non-thermal motions, consistent with feedback-driven expansion phenomena. We find that $\rm P_{TNT}$ is highest in the filamentary region (C4) and lowest at the location of `m4' (C1). This suggests that the contribution of non-thermal pressure is comparatively smaller in the filament than at `m4'. Consequently, the molecular gas within the filament appears to be less disturbed and more dynamically settled. Such conditions are consistent with the filament being in a quiescent state, as previously discussed in this section.

\subsection {Overall physical scenario of the region}

High-resolution CO observations reveal two distinct molecular components in the E71 region. One is associated with the E71 bubble and exhibits signs of feedback-driven expansion, hosting active star-forming features such as a massive arc and the filamentary structure CD. Numerous YSOs are identified in the massive arc with Class\,{\sc i} sources predominantly concentrated along the dust arc, indicating recent star formation activity near the E71 bubble.  In contrast, an adjacent cold filament remains quiescent and devoid of YSOs. Kinematic analysis confirms expanding gas and supersonic turbulence within the bubble, consistent with feedback from a central ionizing source. The stellar cluster Cl1, located inside the bubble, contains a B1.5-type massive star (m2), likely responsible for shaping the bubble through its radiative and mechanical feedback. Multi-wavelength data reveal a PDR, warm dust, and ionized gas, all supporting a triggered star formation scenario. This feedback appears to have initiated the formation of the deeply embedded MYSOs/harbig-haro object, `m4' and `m4$^\prime$', along a surrounding high-extinction arc. We also observe several regularly spaced molecular and dense condensations distributed along the PDR, which serve as strong evidence supporting the ``collect and collapse" mechanism \citep{2003A&A...408L..25D,2015ARep...59..360P,2020ApJ...897...74Z}.

%%%%%%%%%%%%%%%%%%%%%%%%%%%%%%%%%%%%%%%%%%

\section{Summary and Conclusion}\label{sec:conclusion}
We present a comprehensive study of the galactic MIR bubble E71 observed using various telescopes through a multi-wavelength approach. E71 exhibits various encouraging and amusing signatures of star formation activities. We made the following conclusions through our study.

\begin{enumerate}
    
    \item We performed the optical spectroscopy of the probable massive stars `m2' associated with Cl1 using HFOSC and classified it as a B1.5-type star. Thus, we put an upper age limit to the cluster Cl1 as $\sim$12 Myr as the main sequence lifetime of a B1.5-type star. We observed various prominent accretion tracers, i.e., He {\sc i} 1.083 $\mu$m, hydrogen recombination lines from the Brackett and Paschen series, Ca {\sc ii} IRT, Fe {\sc ii} 1.257 $\mu$m in the optical-NIR spectra of the MYSO `m4' obtained through TANSPEC, indicating the accretion processes in `m4'.

    \item We noticed a stellar clustering of slightly elongated morphology (radius $=$ 1.26 pc) inside the E71 bubble. We applied a probabilistic approach on $Gaia$ DR3 PM data to isolate the cluster members and determined the distance of the bubble ($=$ 1.81 $\pm$ 0.15 kpc). We also confirmed this distance through the CMD of the deep-optical IMAGER data. 

    \item The MF slope derived for Cl1 indicates a break around $\sim$0.7 M$_\odot$. In the mass range $\sim0.7< M/M_\odot<5.2$, we determined a slope of $\Gamma = -1.98 \pm 0.33$, while in the lower mass range of $\sim0.2 < M/M_\odot < 0.7$, the slope changes to $\Gamma = +1.58 \pm 0.47$.
    
    \item We also detected a young stellar population showing excess IR emission and performed MST analysis to isolate the cores and the ARs of the cluster associated with the E71 bubble. Since we identified both Class $\textsc{i}$ and Class $\textsc{ii}$ YSOs, we can infer that star formation activities continue to take place surrounding the E71 bubble. Using MST analysis, two main YSO groupings were found: a dense core (within the bubble) and a more extended aggregate region (AR) (outside the bubble). The core includes the Cl1 cluster and overlaps the high-density arc, marking it as an area of intense, ongoing star formation. Cl1 has the lowest Class\,{\sc i} fraction and most compact YSO grouping, suggesting it is older, and the AR shows a slightly higher Class\,{\sc i} fraction, indicating younger stellar content.

    \item We detected an arc-like structure of gas/dust around the massive star `m2' utilizing the \textit{Herschel} MIR/FIR maps. This arc of gas/dust surrounds the diffuse radio emission in the NVSS 1.4 GHz radio continuum map. The \textit{Spitzer} ratio map also indicates the interaction of massive star(s) and their surrounding gas/dust. We also detected a PDR region around the central massive star, evident from the WISE 12 $\mu$m emission, \textit{Spitzer} ratio map (4.5 $\mu$m/3.6 $\mu$m). The MYSO `m4' is found to be located on the southern part of the arc of gas/dust.

    \item The high-resolution uGMRT observation confirms the existence of extended emission in a cavity around massive star `m2' and compact radio emission at the location `m4' and `m4$^\prime$', suggesting the existence of the radio emission, powered by B2-type stars.
       
    \item The E71 bubble is associated with a filamentary molecular cloud in the velocity range [$-20$, $-14$] \kms\, with regularly spaced molecular and dust condensations along its arc. It is possible that `m2' has created a ring of gas and dust, where a new generation of young stars has formed. 

    \item The position-velocity map of \co\, emission suggests that the molecular gas concentrated at the periphery of the E71 bubble, is expanding with an expansion velocity of $\sim$2 \kms. Velocity dispersion and supersonic non-thermal motions confirm expansion dynamics in the bubble, driven by feedback from the central H\,{\sc ii} region.

    \item All these signatures hint toward positive feedback from the massive star `m2' and the possible formation of MYSO `m4' due to it, and suggest that the ``collect and collapse process'' might be a possible model that can describe the ongoing star formation activities around the E71 bubble. 
    
\end{enumerate}
%%%%%%%%%%%%%%%%%%%%%%%%%%%%%%%%%%%%%%%%

%\begin{acknowledgments}
\section*{Acknowledgements}

We thank the anonymous referee for constructive and valuable comments that greatly improved the overall quality of the paper. The observations reported in this paper were obtained by using the 1.3-m DFOT and 3.6-m DOT telescopes at ARIES, Nainital, India, and the 2m HCT at IAO, Hanle, the High Altitude Station of Indian Institute of Astrophysics, Bangalore, India. We thank the staff of the GMRT that made these observations possible. GMRT is run by the National Centre for Radio Astrophysics of the Tata Institute of Fundamental Research. This research used data from the Milky Way Imaging Scroll Painting (MWISP) project, a multi-line survey in 12CO/13CO/C18O along the northern galactic plane with PMO-13.7\,m telescope. We are grateful to all the members of the MWISP working group, particularly the staff members at PMO-13.7\,m telescope, for their long-term support. MWISP was sponsored by the National Key R\&D Program of China with grant 2017YFA0402701 and by CAS Key Research Program of Frontier Sciences with grant QYZDJ-SSW-SLH047.
This publication uses data from the Two Micron All Sky Survey, a joint project of the University of Massachusetts and the Infrared Processing and Analysis Center/California Institute of Technology, funded by the National Aeronautics and Space Administration and the National Science Foundation. This work is based on observations made with the \emph{Spitzer} Space Telescope, operated by the Jet Propulsion Laboratory, California Institute of Technology, under a contract with the National Aeronautics and Space Administration. This publication uses data products from the Wide-field Infrared Survey Explorer, a joint project of the University of California, Los Angeles, and the Jet Propulsion Laboratory/California Institute of Technology, funded by the National Aeronautics and Space Administration. A.V. acknowledges the initial discussion with Ankur Ghosh and Saugata Sarkar on CASA, and the financial support of DST-INSPIRE (No.$\colon$ DST/INSPIRE Fellowship/2019/IF190550). D.K.O. acknowledges the support of the Department of Atomic Energy, Government of India, under project identification No. RTI 4002. R.K.Y. gratefully acknowledges the support from the Fundamental Fund of Thailand Science Research and Innovation (TSRI) (Confirmation No. FFB680072/0269) through the National Astronomical Research Institute of Thailand (Public Organization). A.H. thanks the support by S. N. Bose National Centre for Basic Sciences under the Department of Science and Technology, Govt. of India, and the CSIR-HRDG, Govt. of India for the funding.
The research work at the Physical Research Laboratory is funded by the Department of Space, Government of India. 

%\end{acknowledgments}

% \vspace{5mm}
% \facilities{DFOT:1.3m, DOT:3.6m.}


% \software{APLpy \citep{aplpy2012,aplpy2019}, Astropy \citep{astropy:2013,astropy:2018,astropy:2022}, DAOPHOT-II \citep{1987PASP...99..191S},  
%           Scipy \citep{2020SciPy-NMeth}, Starlink \citep{2014ASPC..485..391C}.
%           }

\appendix

\section{Identification of YSOs in the Region}\label{app:yso_identification}

\begin{figure*}[!ht]
\centering
    \includegraphics[width=0.32\textwidth]{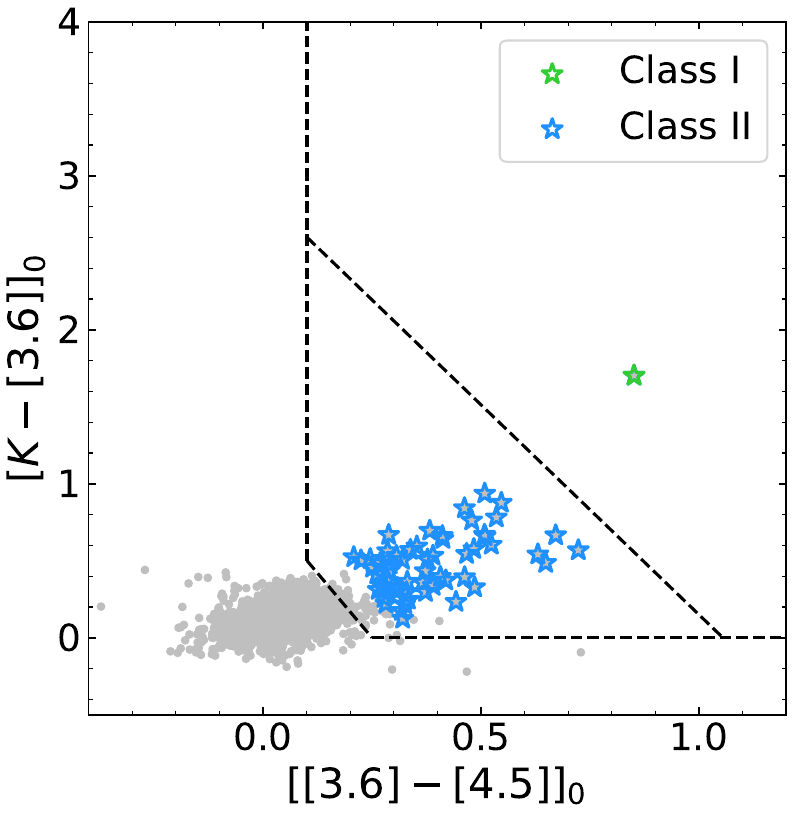}
    \includegraphics[width=0.32\textwidth]{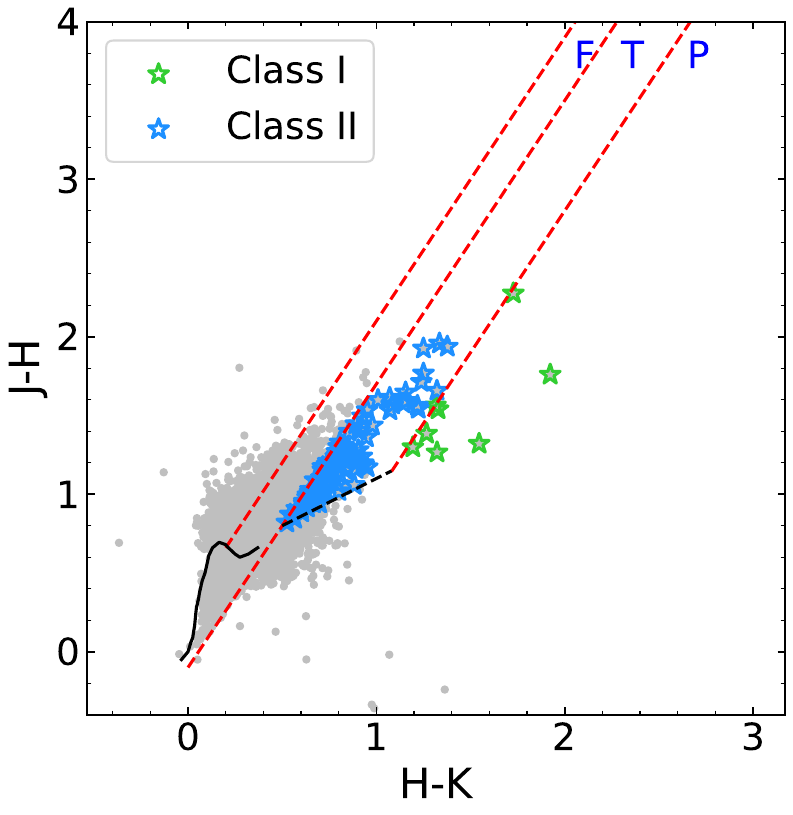}
    \includegraphics[width=0.33\textwidth]{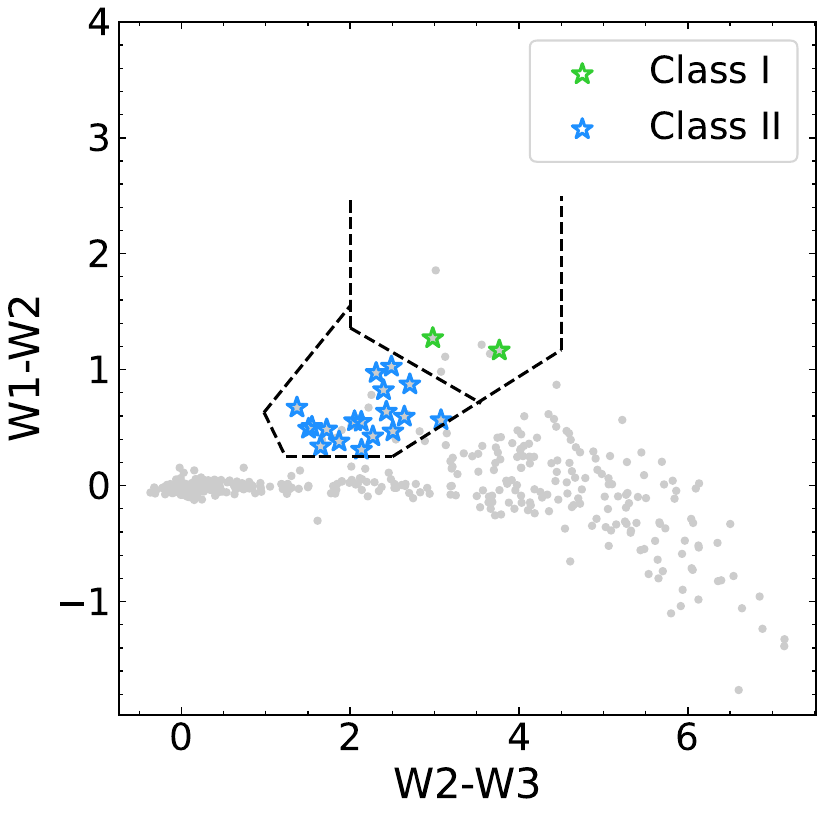}
    \caption{TCD for the YSOs identified within $35^\prime \times35^\prime$ region. Left panel: [K - [3.6]]$_0$ vs. [[3.6] - [4.5]]$_0$ TCD, classification of YSOs is based on the scheme given by \citet{2009ApJS..184...18G}. Middle panel: $(J-H)$ vs. $(H-K)$ CMD, classification is based on the scheme given by \citet{2004ApJ...608..797O}. The parallel dashed lines (red) represent the reddening lines drawn from the tip (spectral type M4) of the giant branch (left red line), from the base (spectral type A0) of the MS branch (middle red line), and the tip of intrinsic CTTS line (right red line). Right Panel: $[W1-W2]$ vs. $[W2-W3]$ TCD, classification of YSOs is based on the scheme given by \citet{2014ApJ...791..131K}. Green and blue asterisks represent Class $\textsc{i}$ and Class $\textsc{ii}$ YSOs, respectively.}
    \label{fig:yso_classification}
\end{figure*}

Identification and classification of YSOs is an essential task in understanding the physical processes occurring in the star-forming regions. YSOs are identified based on their excess IR emission. We used the \emph{GLIMPSE360} catalog of the \emph{Spitzer Science Centre} by employing the classification scheme of \citet{2009ApJS..184...18G}. The [K - [3.6]]$_0$ versus [[3.6] - [4.5]]$_0$ two-color diagram (TCD; see left panel of Figure~\ref{fig:yso_classification}) gives 1 Class $\textsc{i}$ and 57 Class $\textsc{ii}$ YSOs in the target region.

We also used UKIDSS data along with 2MASS data by choosing brighter sources (with \emph{J} magnitude $\leq$13) from 2MASS and the fainter sources (with \emph{J} magnitude $>$13) from UKIDSS. In such a way, we created an NIR catalog. We used the classification scheme of \citet{2004ApJ...608..797O} on this data. The $(J-H)$ versus $(H-K)$ TCD, plotted in the middle panel of Figure~\ref{fig:yso_classification}, gives 8 Class $\textsc{i}$ and 102 Class $\textsc{ii}$ YSOs in the target region. The three parallel lines are drawn from the tip of the giant branch, the base of the main sequence branch, and the tip of intrinsic Classical T Tauri stars (CTTS, these are basically reddening vectors. The sources falling in the ``F" region are either Class $\textsc{iii}$ sources or the field stars; in the ``T" region, either Class $\textsc{ii}$ YSOs or CTSS; and in the ``P" region are classified as Class $\textsc{i}$ YSOs.

Furthermore, we used the MIR data of the ALLWISE catalog of WISE and followed the scheme of \citet{2014ApJ...791..131K}. According to this scheme, a selection criterion is applied on all four WISE bands to first get good quality WISE data, then the extra-galactic contaminants such as AGNs, star-forming galaxies, and transition disks are separated out. We identified 2 Class $\textsc{i}$ and 18 Class $\textsc{ii}$ using ([3.4] - [4.6]) versus ([4.6] - [12]) TCD (see right panel of Figure~\ref{fig:yso_classification}).

We then cross-matched all the YSOs identified using the above three classification schemes with a search radius of 1\arcmin\, and identified 8 Class $\textsc{i}$ and 139 Class $\textsc{ii}$ YSOs in our target region.

\section{Differential Column Density}\label{app:diffcdens}

\begin{figure}[!ht]
    \centering
    \includegraphics[width=\linewidth]{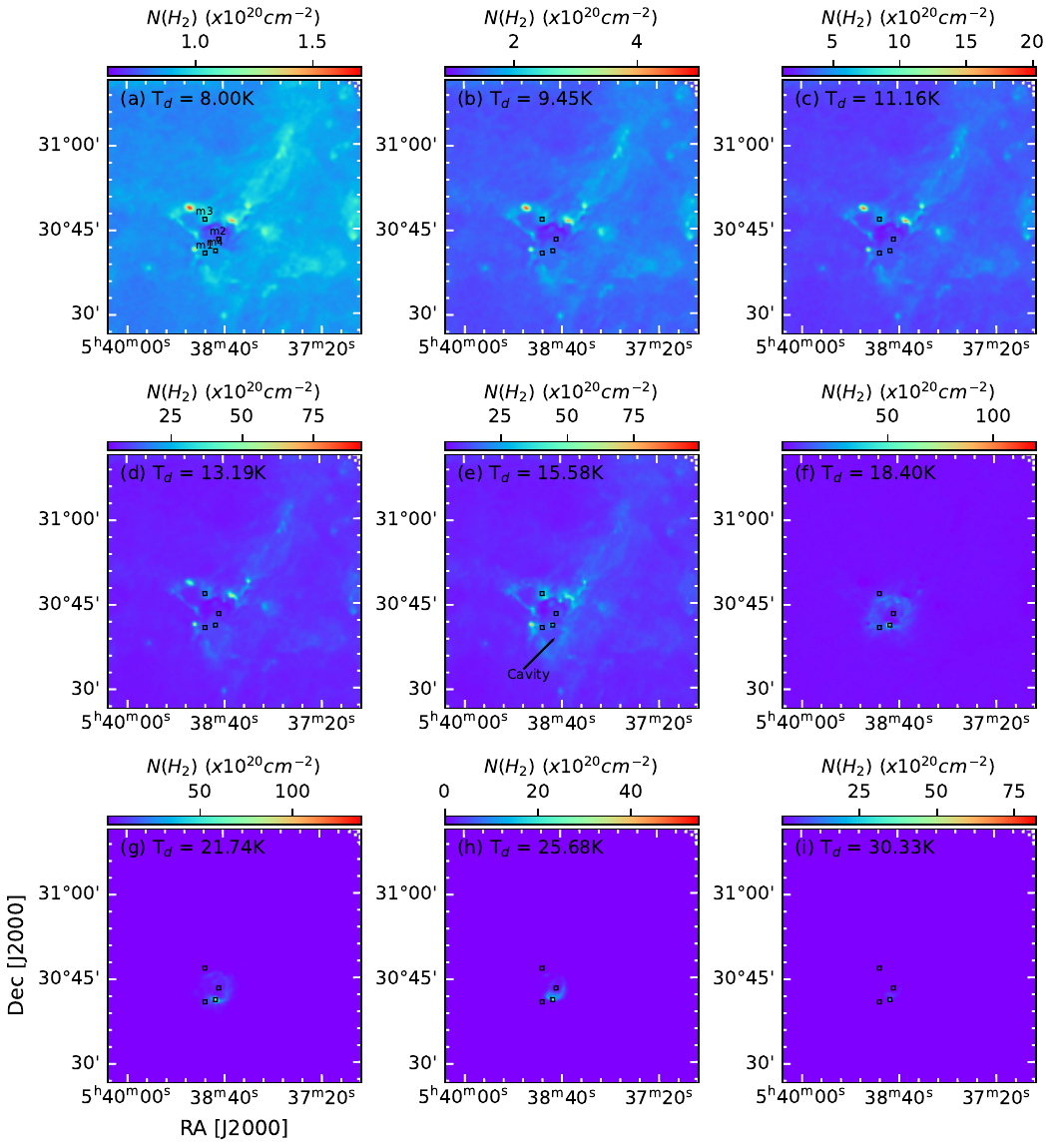}
    \caption{Differential Column Density map at $T_d$, marked in each panel. The locations of `m1, m2, m3, and m4' are also marked in each panel.}
    \label{fig:diffcdens}
\end{figure}

The differential column density provides insight into understanding the distribution of molecular gas with respect to the temperature by tracing different phases of the gas. In Figure~\ref{fig:diffcdens}, we have shown the differential column density for our target region. We observe a filamentary structures up to $T_d = 15.58\,K$; however, at $T_d = 18.40\,K$, it seems like a bubble only. We conclude that the structures below $\sim 16\,K$ are unaffected by the presence of YSOs or protostellar outflows. Additionally, we see a bubble-like structure at $T_d = 15.58\,K$ in the southern direction (marked in Figure~\ref{fig:diffcdens}(e)).
% , which has been marked with yellow color.

\bibliography{e71}{}
\bibliographystyle{aasjournal}

\end{document}